\documentclass[nofootinbib,prd,showpacs]{revtex4}
\usepackage{amsmath}
\usepackage{amsfonts}
\usepackage{amssymb}
\usepackage{graphicx}
\begin{document}
%=================================================================
\title{Generic spherically symmetric dynamic thin-shell traversable wormholes\\ in standard general relativity}
%=================================================================
\author{Nadiezhda Montelongo Garcia}
\email{nmontelongo@fis.cinvestav.mx }
\affiliation{Centro de Astronomia
e Astrof\'{\i}sica da Universidade de Lisboa, Campo Grande, Edif\'{i}cio C8
1749-016 Lisboa, Portugal}
\affiliation{Departamento de F\'isica, Centro de Investigaci\'on 
y Estudios avanzados del I.P.N.,\\ A.P. 14-700,07000 M\'exico, DF, M\'exico}
%=================================================================
\author{Francisco S.~N.~Lobo}
\email{flobo@cii.fc.ul.pt}\affiliation{Centro de Astronomia
e Astrof\'{\i}sica da Universidade de Lisboa, Campo Grande, Edif\'{i}cio C8
1749-016 Lisboa, Portugal}
%=================================================================
\author{Matt Visser}
\email{matt.visser@msor.vuw.ac.nz}\affiliation{School of Mathematics, Statistics, and Operations
Research, \\Victoria University of Wellington, P.O. Box 600, Wellington 6140, New Zealand}
%%%%%%%%%%%%%%%%%%%%%%%%%%%%%%%%%%%%%%%%%%%%%%%%%%%%%%%%%%%%%%%%
\date{9 December 2011; \LaTeX-ed \today}
%=================================================================
\begin{abstract}
%=================================================================
We consider the construction of generic spherically symmetric thin-shell traversable wormhole spacetimes in standard general relativity. By using the cut-and-paste procedure, we comprehensively analyze the stability of arbitrary spherically symmetric thin-shell wormholes to linearized spherically symmetric perturbations around static solutions. While a number of special cases have previously been dealt with in scattered parts of the literature, herein we take considerable effort to make the analysis as general and unified as practicable.
We demonstrate in full generality that stability of the wormhole is equivalent to choosing suitable properties for the exotic material residing on the wormhole throat.
%=================================================================
\end{abstract}
%=================================================================
%=================================================================
\maketitle
%%%%%%%%%%%%%%%%%%%%%%%%%%%%%%%%%%%%%%%%%%%%%%%%%%%%%%%%%%%%%%%%
\def\d{{\mathrm{d}}}
%%%%%%%%%%%%%%%%%%%%%%%%%%%%%%%%%%%%%%%%%%%%%%%%%%%%%%%%%%%%%%%%
%%%%%%%%%%%%%%%%%%%%%%%%%%%%%%%%%%%%%%%%%%%%%%%%%%%%%%%%%%%%%%%%
\section{Introduction}
%%%%%%%%%%%%%%%%%%%%%%%%%%%%%%%%%%%%%%%%%%%%%%%%%%%%%%%%%%%%%%%%

Traversable wormholes are hypothetical tunnels in spacetime, through which observers may
freely travel~\cite{Morris, 269623}. These geometries are supported by ``exotic matter'', involving a
stress-energy tensor violating the null energy condition (NEC).  That is, there exists at
least one null vector $k^\mu$ such that  $T_{\mu\nu}k^{\mu}k^{\nu}<0$ on or in the immediate
vicinity of the wormhole throat (see, in particular,~\cite{Visser, gr-qc/9704082, gr-qc/9710001}). In fact, wormhole geometries violate all the standard
pointwise energy conditions, and all the standard averaged energy conditions~\cite{Visser}.
Although (most) classical forms of matter are believed to obey (most of) the standard energy
conditions~\cite{hawkingellis}, it is a well-known fact that they are violated by
certain quantum effects, amongst which we may refer to the Casimir effect, Hawking
evaporation, and vacuum polarization. (See~\cite{Visser}, and more recently~\cite{gr-qc/0205066}, for a review. For some technical details see~\cite{gvp}.) 
It is interesting to note that the known
violations of the pointwise energy conditions led researchers to consider the possibility of
averaging of the energy conditions over timelike or null geodesics~\cite{Tipler-AEC}. For
instance, the averaged weak energy condition (AWEC) states that the integral
of the energy density measured by a geodesic observer is non-negative. (That is, $\int
T_{\mu\nu}U^\mu U^\nu \,d\tau \geq 0$, where $\tau$ is the observer's proper time.) Thus, the
averaged energy conditions are weaker than the pointwise energy conditions, they permit
localized violations of the energy conditions, as long as they hold when suitably averaged
along a null or timelike geodesic~\cite{Tipler-AEC}.

As the theme of exotic matter is a problematic issue, it is useful to minimize its usage.
In fact, it is important to emphasize that the theorems which guarantee the energy condition
violation are remarkably silent when it comes to making quantitative statements regarding the
``total amount'' of energy condition violating matter in the spacetime. In this context, a
suitable measure for quantifying this notion was developed in \cite{VKD}, see also~\cite{gr-qc/0405103}, where it was shown that wormhole geometries may in principle be supported by arbitrarily small quantities of exotic matter (an interesting application of the quantification of the total amount of energy condition violating matter in warp drive spacetimes was considered in \cite{gr-qc/0406083}). In the context of minimizing the usage of exotic matter, it was also found
that for specific models of stationary and axially symmetric traversable wormholes the exotic
matter is confined to certain regions around the wormhole throat, so that certain
classes of geodesics traversing the wormhole need not encounter any energy condition violating
matter \cite{Teo:1998dp}. For dynamic wormholes the null energy condition, more precisely the
averaged null energy condition, can be avoided in certain regions \cite{hochvisserPRL98,
hochvisserPRD98, gr-qc/9901020}. Evolving wormhole geometries were also found which exhibit ``flashes'' of
weak energy condition (WEC) violation, where the matter threading the wormhole violates the
energy conditions for small intervals of time \cite{Kar}. In the context of nonlinear
electrodynamics, it was found that certain dynamic wormhole solutions obey (suitably defined
versions of) the WEC \cite{Arellano:2006ex}.

It is interesting to note that in modified theories of gravity, more specifically in $f(R)$
gravity, the matter threading the wormhole throat can be forced to obey (suitably defined
versions of) all the energy conditions, and it is the higher-order curvature terms that are
responsible for supporting these wormhole geometries \cite{modgravity}. (Related issues arise when scalar fields are conformally coupled to gravity~\cite{gr-qc/0003025, gr-qc/9908029}.) In braneworlds, it was
found that it is a combination of the local high-energy bulk effects, and the nonlocal
corrections from the Weyl curvature in the bulk, that may induce an effective NEC violating
signature on the brane, while the ``physical''/``observable'' stress-energy tensor confined to
the brane, threading the wormhole throat, nevertheless satisfies the energy
conditions~\cite{Lobo:2007qi}.

An interesting and efficient manner to minimize the violation of the null energy condition,
extensively analyzed in the literature, is to construct thin-shell wormholes using the 
thin-shell formalism \cite{Visser,thinform,Israel} and the cut-and-paste procedure as described
in~\cite{Visser,VisserPRD,VisserNPB,VisserPLB,Poisson}. Motivated in minimizing
the usage of exotic matter, the thin-shell construction was generalized to nonspherically
symmetric cases~\cite{Visser,VisserPRD}, and in particular, it was found that a traveler may
traverse through such a wormhole without encountering regions of exotic matter. In the context
of a (limited) stability analysis, in \cite{VisserNPB}, two Schwarzschild spacetimes were surgically
grafted together in such a way that no event horizon is permitted to form. This surgery
concentrates a nonzero stress energy on the boundary layer between the two asymptotically
flat regions and a dynamical stability analysis {(with respect to spherically symmetric perturbations)} was explored. In the latter stability
analysis, constraints were found on the equation of state of the exotic matter that comprises
the throat of the wormhole. Indeed, the stability of the latter thin-shell wormholes was
considered for certain specially chosen equations of state \cite{Visser,VisserNPB}, where the
analysis addressed the issue of stability in the sense of proving bounded motion for the
wormhole throat. This dynamical analysis was generalized to the stability of spherically
symmetric thin-shell wormholes by considering linearized radial perturbations around some
assumed static solution of the Einstein field equations, without the need to specify an
equation of state \cite{Poisson}. This linearized stability analysis around a static solution
was soon generalized to the presence of charge \cite{Eiroa}, and of a cosmological constant
\cite{LoboCraw}, and was subsequently extended to a plethora of individual scenarios \cite{thinshellwh,Ishak}, some of them rather ad hoc.

The key point of the present paper is to develop an extremely general, flexible, and robust framework that can
quickly be adapted to general spherically symmetric traversable wormholes in 3+1 dimensions. We shall consider
standard general relativity, with traversable wormholes that are spherically symmetric, with
all of the exotic material confined to a thin shell. The bulk spacetimes on either side of the
wormhole throat will be spherically symmetric and static but otherwise arbitrary (so the
formalism is simultaneously capable of dealing with wormholes embedded in Schwarzschild,
Reissner--Nordstr\"om, Kottler, or de~Sitter spacetimes, or even ``stringy'' black hole
spacetimes).  The thin shell (wormhole throat), while constrained by spherical symmetry,  will otherwise be permitted to move freely in the bulk
spacetimes, permitting a fully dynamic analysis. This will then allow us to perform a general
stability analysis against spherically symmetric perturbations, where wormhole stability is related to the properties of the exotic matter
residing on the wormhole throat. We particularly emphasize that our analysis can deal with geometrically imposed ``external forces'', (to be more fully explained below), a feature that has so far been missing from the published literature. Additionally we emphasize the derivation of rather explicit and very general rules relating the internal structure of the wormhole throat to a ``potential'' that drives the motion of the throat.

This paper is organized in the following manner: In Sec. \ref{secII} we outline in detail
the general formalism of generic dynamic spherically symmetric thin-shell wormholes, and
provide a novel approach to the linearized stability analysis around a static solution. In
Sec. \ref{secIII}, we provide specific examples by applying the generic linearized
stability formalism outlined in the previous section. In Sec. \ref{conclusion}, we draw some general  conclusions.

%###########################################################
\section{Formalism}
%###########################################################
\label{secII}
%###########################################################

Let us first perform a general theoretical analysis; subsequently we shall look at a number of specific examples.

%###########################################################
\subsection{Bulk spacetimes}
%###########################################################

To set the stage, consider two distinct spacetime manifolds, ${\cal M_+}$ and ${\cal M_-}$,
with metrics given by $g_{\mu \nu}^+(x^{\mu}_+)$ and $g_{\mu \nu}^-(x^{\mu}_-)$, in terms 
of independently defined coordinate systems $x^{\mu}_+$ and $x^{\mu}_-$. A single manifold 
${\cal M}$ is obtained by gluing together the two distinct manifolds, ${\cal M_+}$ and 
${\cal M_-}$, i.e., ${\cal M}={\cal M_+}\cup {\cal M_-}$, at their boundaries. The latter 
are given by $\Sigma_+$ and $\Sigma_-$, respectively, with the natural identification of 
the boundaries $\Sigma=\Sigma_+=\Sigma_-$. This construction is depicted in the embedding diagram in Fig. \ref{fig1}, and is further explored in the sections given below. 
\begin{figure}[!htb]
  %\centering
  \includegraphics[width=3.5 in]{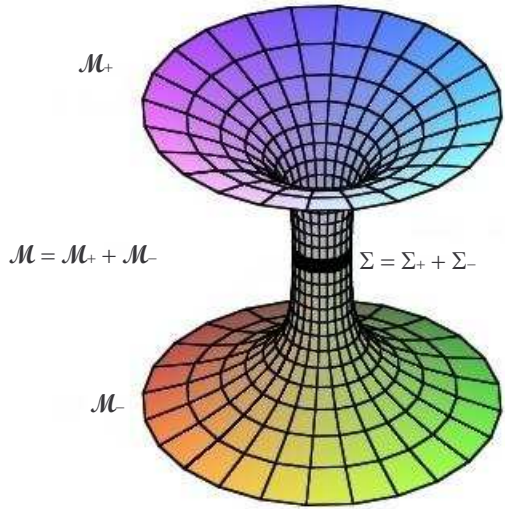}
  \caption{The figure depicts an embedding diagram of a traversable thin-shell wormhole.
  A single manifold ${\cal M}$ is obtained by gluing together two distinct spacetime 
  manifolds, ${\cal M_+}$ and ${\cal M_-}$, 
  at their boundaries, i.e., ${\cal M}={\cal M_+}\cup {\cal M_-}$, with the natural 
  identification of the boundaries $\Sigma=\Sigma_+=\Sigma_-$.
  }
  \label{fig1}
\end{figure}

More specifically, throughout this work, we consider two generic static spherically 
symmetric spacetimes given by the following line elements:
\begin{eqnarray}
ds^2 = - e^{2\Phi_{\pm}(r_{\pm})}\left[1-\frac{b_{\pm}(r_{\pm})}{r_{\pm}}\right] dt_{\pm}^2 
+
\left[1-\frac{b_{\pm}(r_{\pm})}{r_{\pm}}\right]^{-1}\,dr_{\pm}^2 + r_{\pm}^2 
d\Omega_{\pm}^{2},
\label{generalmetric}
\end{eqnarray}
on ${\cal M_\pm}$, respectively.
Using the Einstein field equation, $G_{{\mu}{\nu}}=8\pi \,T_{{\mu}{\nu}}$ (with $c=G=1$), the
(orthonormal) stress-energy tensor components in the bulk are given by
\begin{eqnarray}
\rho(r)&=&\frac{1}{8\pi r^{2}}b',\label{rho}\\
p_{r}(r)&=&-\frac{1}{8\pi r^{2}}\left[2\Phi '(b-r)+b'\right],\label{pr}\\
p_{t}(r)&=&-\frac{1}{16\pi r^{2}}[(-b+3rb'-2r)\Phi '+2r(b-r)(\Phi ')^{2}+2r(b-r)\Phi ''
+b''r]\,,
\label{pt}
\end{eqnarray}
where the prime denotes a derivative with respect to the radial coordinate. Here $\rho(r)$ is the
energy density, $p_r(r)$ is the radial pressure, and $p_t(r)$ is the lateral pressure measured in
the orthogonal direction to the radial direction. The $\pm$ subscripts were (temporarily) dropped so
as not to overload the notation.

The energy conditions will play an important role in the analysis that follows, so we will at this
stage define the NEC. The latter is satisfied if $T_{\mu\nu}\,k^\mu\,k^\nu
\geq 0$, where $T_{\mu\nu}$ is the stress-energy tensor and $k^{\mu}$ any null vector. Along the
radial direction, with $k^{\hat{\mu}}=(1,\pm 1,0,0)$ in the orthonormal frame, where
$T_{\hat{\mu}\hat{\nu}}={\rm diag}[\rho(r),p_r(r),p_t(r),p_t(r)]$, we then have the particularly simple condition
\begin{equation}
 T_{\hat{\mu}\hat{\nu}}\,k^{\hat{\mu}}\,k^{\hat{\nu}}=\rho(r)+p_r(r)
 =\frac{(r-b)\Phi^{'}}{4\pi r^{2}}\geq 0.    \label{generalNEC}
\end{equation}
Note that in any region where the  $t$ coordinate is timelike (requiring $r>b(r)$) the radial NEC
reduces to $\Phi'(r)>0$. The NEC in the transverse direction, $\rho+p_t\geq 0$, does not have any
direct simple interpretation in terms of the metric components.

\begin{figure}[!htb]
  %\centering
  \includegraphics[width=2.5 in]{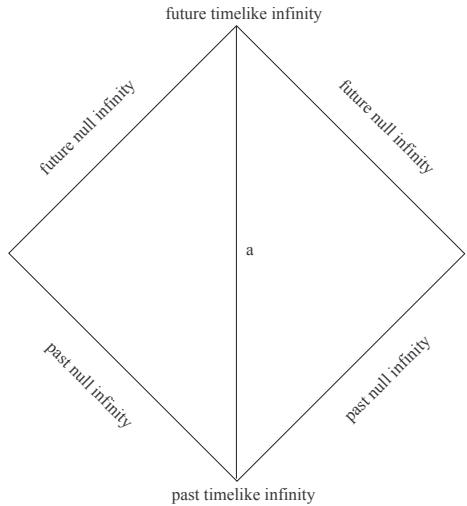}
  \caption{The spacetime diagram for a thin-shell wormhole in an asymptotically flat 
  spacetime, represented by two manifolds ${\cal M_+}$ and ${\cal 
  M_-}$, joined at a radial coordinate $a$, at the junction surface $\Sigma$.
  }
  \label{fig3}
\end{figure}

We emphasize that the results outlined in this work are also valid for the case of an intra-universe wormhole, i.e., with a single manifold with a wormhole connecting distant regions. This is indeed true as long as the bulk geometries are both asymptotically flat, then they can be viewed as widely separated parts of the same asymptotically flat spacetime (to an arbitrarily good approximation that gets better as the two wormhole mouths get further and further separated). For instance, see Fig. \ref{fig3} for a spacetime diagram depicting a thin-shell wormhole in an asymptotically flat spacetime, represented by two manifolds joined at a radial coordinate $a$,

%###########################################################
\subsection{Extrinsic curvature}
%###########################################################

The manifolds are bounded by hypersurfaces $\Sigma_+$ and $\Sigma_-$, respectively, with induced
metrics $g_{ij}^+$ and $g_{ij}^-$. The hypersurfaces are isometric, i.e.,
$g_{ij}^+(\xi)=g_{ij}^-(\xi)=g_{ij}(\xi)$, in terms of the intrinsic coordinates, invariant under
the isometry. As mentioned above, single manifold ${\cal M}$ is obtained by gluing together ${\cal
M_+}$ and ${\cal M_-}$ at their boundaries, i.e., ${\cal M}={\cal M_+}\cup {\cal M_-}$, with the
natural identification of the boundaries $\Sigma=\Sigma_+=\Sigma_-$. The three holonomic basis
vectors ${\bf e}_{(i)}=\partial /\partial \xi^i$ tangent to $\Sigma$ have the following components
$e^{\mu}_{(i)}|_{\pm}=\partial x_{\pm}^{\mu}/\partial \xi^i$, which provide the induced metric on
the junction surface by the following scalar product $g_{ij}={\bf e}_{(i)}\cdot {\bf e}_{(j)}=g_{\mu
\nu}e^{\mu}_{(i)}e^{\nu}_{(j)}|_{\pm}$. The intrinsic metric to $\Sigma$ is thus provided by
\begin{equation}
ds^2_{\Sigma}=-d\tau^2 + a(\tau)^2 \,(d\theta ^2+\sin
^2{\theta}\,d\phi^2).
\end{equation}

Thus, for the static and spherically symmetric spacetime considered in this work, the single
manifold, ${\cal M}$, is obtained by gluing ${\cal M_+}$ and ${\cal M_-}$ at $\Sigma$, i.e., at
$f(r,\tau)=r-a(\tau)=0$. The position of the junction surface is given by $x^{\mu}(\tau,\theta,
\phi)=(t(\tau),a(\tau),\theta,\phi)$, and the respective $4$-velocities (as measured in the static
coordinate systems on the two sides of the junction) are
\begin{eqnarray}
U^{\mu}_{\pm}=
\left(\frac{e^{-\Phi_{\pm}(a)}\sqrt{1-\frac{b_{\pm}(a)}{a}+\dot{a}^{2}}}{1-\frac{b_{\pm}(a)}{a}},\;
\dot{a},0,0 \right),
\end{eqnarray}
where the overdot denotes a derivative with respect to $\tau$, which is the proper time of an
observer comoving with the junction surface.

We shall consider a timelike junction surface $\Sigma$, defined by the parametric equation of the
form $f(x^{\mu}(\xi^i))=0$. The unit normal $4-$vector, $n^{\mu}$, to $\Sigma$ is defined as
\begin{equation}\label{defnormal}
n_{\mu}=\pm \,\left |g^{\alpha \beta}\,\frac{\partial f}{\partial
x ^{\alpha}} \, \frac{\partial f}{\partial x ^{\beta}}\right
|^{-1/2}\;\frac{\partial f}{\partial x^{\mu}}\,,
\end{equation}
with $n_{\mu}\,n^{\mu}=+1$ and $n_{\mu}e^{\mu}_{(i)}=0$. The Israel formalism requires that the
normals point from ${\cal M_-}$ to ${\cal M_+}$ \cite{Israel}. Thus, the unit normals to the
junction surface, determined by Eq.~(\ref{defnormal}), are given by
\begin{eqnarray}
n^{\mu}_{\pm}= \pm \left(
\frac{e^{-\Phi_{\pm}(a)}}{1-\frac{b_{\pm}(a)}{a}}\;\dot{a},
\sqrt{1-\frac{b_{\pm}(a)}{a}+\dot{a}^2},0,0
\right) \label{normal}
\,.
\end{eqnarray}
Note (in view of the spherical symmetry) that the above expressions can also be deduced from the contractions $U^{\mu}n_{\mu}=0$ and $n^{\mu}n_{\mu}=+1$.
The extrinsic curvature, or the second fundamental form, is defined as $K_{ij}=n_{\mu;
\nu}e^{\mu}_{(i)}e^{\nu}_{(j)}$. Differentiating $n_{\mu}e^{\mu}_{(i)}=0$ with respect to $\xi^j$,
we have
\begin{equation}
n_{\mu}\frac{\partial ^2 x^{\mu}}{\partial \xi^i \, \partial \xi^j}=
-n_{\mu,\nu}\, \frac{\partial x^{\mu}}{\partial \xi^i}\frac{\partial x^{\nu}}{\partial \xi^j},
\end{equation}
so that in general the extrinsic curvature is given by
\begin{eqnarray}
\label{extrinsiccurv}
K_{ij}^{\pm}=-n_{\mu} \left(\frac{\partial ^2 x^{\mu}}{\partial
\xi ^{i}\,\partial \xi ^{j}}+\Gamma ^{\mu \pm}_{\;\;\alpha
\beta}\;\frac{\partial x^{\alpha}}{\partial \xi ^{i}} \,
\frac{\partial x^{\beta}}{\partial \xi ^{j}} \right) \,.
\end{eqnarray}
Note that for the case of a thin shell $K_{ij}$ is not continuous across $\Sigma$, so that for
notational convenience, the discontinuity in the second fundamental form is defined as
$\kappa_{ij}=K_{ij}^{+}-K_{ij}^{-}$.

Using equation (\ref{extrinsiccurv}), the non-trivial components of the extrinsic curvature can
easily be computed to be
\begin{eqnarray}
K^{\theta \;\pm }_{\;\;\theta}&=&\pm\frac{1}{a}\,\sqrt{1-\frac{b_{\pm}(a)}{a}+\dot{a}^2}\;,
\label{genKplustheta}
\\
K^{\tau\;\pm}_{\;\;\tau}&=&\pm\,
\left\{\frac{\ddot a+\frac{b_{\pm}(a)-b'_{\pm}(a)a}{2a^2}}{\sqrt{1-\frac{b_{\pm}(a)}{a}+\dot{a}^{2}}}
+ \Phi'_{\pm}(a) \sqrt{1-\frac{b_{\pm}(a)}{a}+\dot{a}^{2}} \right\}  \,,
\label{genKminustautau}
\end{eqnarray}
where the prime now denotes a derivative with respect to the coordinate $a$.

Several comments are in order:
\begin{itemize}

\item
Note that $K^{\theta \;\pm }_{\;\;\theta}$ is independent of the quantities $\Phi_\pm$; a circumstance that will
have implications later in this article. This is most easily verified by noting that in terms of the
normal distance $\ell$ to the shell $\Sigma$ the extrinsic curvature can be written as $K_{ij} =
{1\over2} \partial_\ell g_{ij} = {1\over2} n^\mu \partial_\mu g_{ij} =  {1\over2} n^r \partial_r
g_{ij}$, where the last step relies on the fact that the bulk spacetimes are static. Then since
$g_{\theta\theta}=r^2$, differentiating and setting $r\to a$ we have $K_{\theta\theta} = a \; n^r$.
Finally
\begin{equation}
K^{\theta \;\pm }_{\;\;\theta} = {n^r\over a},
\end{equation}
which is a particularly simple formula in terms of the radial component of the normal vector, and
which easily lets us verify \eqref{genKplustheta}.

\item
For $K_{\tau\tau}$ there is an argument (easily extendable to the present context) in
Ref.~\cite{Visser} (see especially pages 181--183) to the effect that
\begin{equation}
K^{\tau\;\pm}_{\;\;\tau}=\pm\, \hbox{(magnitude of the physical 4-acceleration of the throat)}.
\end{equation}
This gives a clear physical interpretation to $K^{\tau\;\pm}_{\;\;\tau}$ and rapidly allows one to
verify \eqref{genKminustautau}.

\item
There is also an important differential relationship between these extrinsic curvature components
\begin{equation}
{\d\over\d\tau}\left\{ a \; e^{\Phi_\pm} \; K^{\theta \;\pm }_{\;\;\theta} \right\} = e^{\Phi_\pm}
\; K^{\tau\;\pm}_{\;\;\tau} \; \dot a.
\label{E:differential}
\end{equation}
The most direct way to verify this is to simply differentiate, using \eqref{genKplustheta} and
\eqref{genKminustautau} above. Geometrically, the existence of these relations between the extrinsic
curvature components is ultimately due to the fact that the bulk spacetimes have been chosen to be
static. By noting that
\begin{equation}
{\d\over \d a}\left({1\over2} \dot a^2\right) = \left({\d\over \d a} \dot a\right)  \dot a = \ddot a,
\label{E:trick}
\end{equation}
we can also write this differential relation as
\begin{equation}
{\d\over\d a}\left\{ a \; e^{\Phi_\pm} \; K^{\theta \;\pm }_{\;\;\theta} \right\} = e^{\Phi_\pm}
\; K^{\tau\;\pm}_{\;\;\tau}.
\end{equation}

\item
We emphasize that including the possibility that  $\Phi_\pm(a)\neq 0$ is already a significant generalization of the extant literature.
 \end{itemize}

%###########################################################
\subsection{Lanczos equations: Surface stress-energy}
%###########################################################

\begin{figure}[!htb]
  %\centering
  \includegraphics[width=3.5 in]{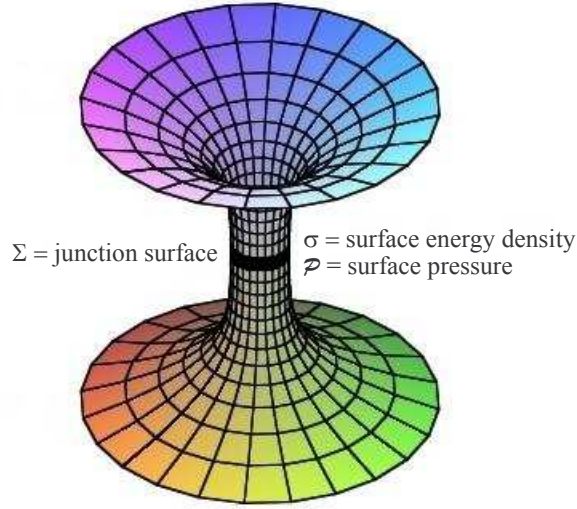}
  \caption{The figure depicts an embedding diagram of a traversable thin-shell wormhole.
  The single manifold ${\cal M}$ is obtained by gluing together ${\cal M_+}$ and ${\cal 
  M_-}$, at junction surface $\Sigma$.  The surface stress-energy tensor on $\Sigma$ is 
  given  in terms of the surface energy density, $\sigma$, and the surface pressure, ${\cal 
  P}$. See the text for details.
  }
  \label{fig2}
\end{figure}

The Lanczos equations follow from the Einstein equations applied to the hypersurface joining the
bulk spacetimes, and are given by
\begin{equation}
S^{i}_{\;j}=-\frac{1}{8\pi}\,(\kappa ^{i}_{\;j}-\delta
^{i}_{\;j}\kappa ^{k}_{\;k})  \,.
\end{equation}
Here $S^{i}_{\;j}$ is the surface stress-energy tensor on $\Sigma$. In particular, because 
of spherical symmetry considerable simplifications occur, namely $\kappa ^{i}_{\;j}={\rm 
diag}\left(\kappa ^{\tau}_{\;\tau},\kappa ^{\theta}_{\;\theta},\kappa^{\theta}_{\;
\theta}\right)$. The surface stress-energy tensor may be written in terms of the surface 
energy density, $\sigma$, and the surface pressure, ${\cal P}$, as $S^{i}_{\;j}={\rm diag}
(-\sigma,{\cal P},{\cal P})$. The Lanczos equations then reduce to
\begin{eqnarray}
\sigma &=&-\frac{1}{4\pi}\,\kappa ^{\theta}_{\;\theta} \,,\label{sigma} \\
{\cal P} &=&\frac{1}{8\pi}(\kappa ^{\tau}_{\;\tau}+\kappa
^{\theta}_{\;\theta}) \,. \label{surfacepressure}
\end{eqnarray}
Taking into account the computed extrinsic curvatures,
Eqs.~(\ref{genKplustheta})--(\ref{genKminustautau}), we see that
Eqs.~(\ref{sigma})--(\ref{surfacepressure}) provide us with the following expressions for the
surface stresses:
\begin{eqnarray}
\sigma&=&-\frac{1}{4\pi a}\left[\sqrt{1-\frac{b_{+}(a)}{a}
+\dot{a}^{2}}+\sqrt{1-\frac{b_{-}(a)}{a}+\dot{a}^{2}}\,\right],
\label{gen-surfenergy2}
\\
{\cal P}&=&\frac{1}{8\pi a}\left[
\frac{1+\dot{a}^2+a\ddot{a}-\frac{b_+(a)+ab'_+(a)}{2a}}{\sqrt{1-\frac{b_{+}(a)}{a}+\dot{a}^{2}}}
+
\sqrt{1-\frac{b_{+}(a)}{a}+\dot{a}^{2}} \; a\Phi'_{+}(a)
\right. 
\nonumber\\
&&
\qquad
\left. +
\frac{1+\dot{a}^2+a\ddot{a}-\frac{b_-(a)+ab'_-(a)}{2a}}{\sqrt{1-\frac{b_{-}(a)}{a}+\dot{a}^{2}}}
+
\sqrt{1-\frac{b_{-}(a)}{a}+\dot{a}^{2}} \; a\Phi'_{-}(a)
\right].
\label{gen-surfpressure2}
\end{eqnarray}
The surface stress-energy tensor on the junction surface $\Sigma$ is depicted in the 
embedding spacetime diagram in Fig. \ref{fig2}, in terms of the surface energy density, 
$\sigma$, and the surface pressure, ${\cal P}$.

Note that the surface mass of the thin shell is given by $m_s=4\pi a^2\sigma$, a quantity 
which will be extensively used below.  Note further that the surface energy density 
$\sigma$ is always negative, (which is where the energy condition violations show up in 
this thin-shell context), and furthermore that the surface energy density $\sigma$ is 
independent of the two quantities $\Phi_\pm$.

%##########################################################################
\subsection{Conservation identity}
%##########################################################################

The first contracted Gauss--Codazzi equation, sometimes called simply the Gauss equation,  or in
general relativity more often referred to as the ``Hamiltonian constraint'', is
\begin{eqnarray}
G_{\mu \nu}\;n^{\mu}\,n^{\nu}=\frac{1}{2}\,(K^2-K_{ij}K^{ij}-\,^3R)\,.
    \label{1Gauss}
\end{eqnarray}
Together with the Einstein equations this provides the evolution identity
\begin{eqnarray}
S^{ij}\;\overline{K}_{ij}=-\left[T_{\mu \nu}n^{\mu}n^{\nu}
\right]^{+}_{-}\,.
\end{eqnarray}
The convention $\left[X \right]^+_-\equiv
X^+|_{\Sigma}-X^-|_{\Sigma}$ and $\overline{X} \equiv {1\over2}
(X^+|_{\Sigma}+X^-|_{\Sigma})$ is used.
The second contracted Gauss--Codazzi equation, sometimes called simply the Codazzi or the Codazzi--Mainardi equation, or in general relativity more often referred to as the ``ADM constraint'' or
``momentum constraint'', is
\begin{eqnarray}
G_{\mu \nu}e^{\mu}_{(i)}n^{\nu}=K^j_{i|j}-K,_{i}\,.
    \label{2Gauss}
\end{eqnarray}
Together with the Lanczos equations this provides the conservation identity
\begin{eqnarray}\label{conservation}
S^{i}_{\;j|i}=\left[T_{\mu \nu}\; e^{\mu}_{\;(j)}n^{\nu}\right]^+_-\,.
\end{eqnarray}
When interpreting the conservation identity Eq.~(\ref{conservation}), consider first the momentum
flux defined by
\begin{equation}
\left[T_{\mu\nu}\; e^{\mu}_{\;(\tau)}\,n^{\nu}\right]^+_-
=\left[T_{\mu\nu}\; U^{\mu}\,n^{\nu}\right]^+_-
=\left[\pm
\left(T_{\hat{t}\hat{t}}+T_{\hat{r}\hat{r}}\right)
\,\frac{\dot{a}\sqrt{1-\frac{b(a)}{a}+\dot{a}^{2}}}{1-\frac{b(a)}{a}} \;
\right]^+_-\,,
\label{flux}
\end{equation}
where $T_{\hat{t}\hat{t}}$ and $T_{\hat{r}\hat{r}}$ are the bulk stress-energy tensor components
given in an orthonormal basis. This flux term corresponds to the net discontinuity in the (bulk)
momentum flux $F_\mu=T_{\mu\nu}\,U^\nu$ which impinges on the shell. (This flux term is identically zero in all the currently extant literature.) Applying the (bulk) Einstein
equations we see
\begin{equation}
\left[T_{\mu\nu}\; e^{\mu}_{\;(\tau)}\,n^{\nu}\right]^+_-
=\frac{\dot a}{4\pi a}\, \left[\Phi_+'(a)\sqrt{1-\frac{b_+(a)}{a}+\dot{a}^{2}}
+ \Phi_-'(a)\sqrt{1-\frac{b_-(a)}{a}+\dot{a}^{2}}\,\right]\,.
\end{equation}
It is useful to define the quantity
\begin{equation}
\Xi
=\frac{1}{4\pi a}\, \left[\Phi_+'(a)\sqrt{1-\frac{b_+(a)}{a}+\dot{a}^{2}}
+ \Phi_-'(a)\sqrt{1-\frac{b_-(a)}{a}+\dot{a}^{2}}\,\right]\,.
\end{equation}
and to let $A=4\pi a^2$ be the surface area of the thin shell. Then in the general case, the
conservation identity provides the following relationship
\begin{equation}
\frac{d\sigma}{d\tau}+(\sigma+{\cal P})\,{1\over A} \frac{dA}{d\tau}=\Xi \, \dot{a}\,,
\label{E:conservation2}
\end{equation}
or equivalently
\begin{equation}
\frac{d(\sigma A)}{d\tau}+{\cal P}\,\frac{dA}{d\tau}=\Xi \,A \, \dot{a}\,,
\label{E:conservation3}
\end{equation}
The first term represents the variation of the internal energy of the shell, the second term is the
work done by the shell's internal force, and the third term represents the work done by the external
forces.  Once could also brute force verify this equation by explicitly differentiating \eqref{gen-surfenergy2} using \eqref{gen-surfpressure2} and the relations \eqref{E:differential}.
If we assume that the equations of motion can be integrated to determine the surface energy
density as a function of radius $a$, that is, assuming the existence of a suitable function
$\sigma(a)$, then the conservation equation can be written as
\begin{equation}
\sigma'=-\frac{2}{a}\,(\sigma +{\cal P})+\Xi\,,
\label{cons-equation}
\end{equation}
where $ \sigma'=d\sigma /da$. We shall carefully analyze the integrability conditions for
$\sigma(a)$ in the next sub-section. For now, note that the flux term (external force term) $\Xi$ is
automatically zero whenever $\Phi_\pm=0$; this is actually a quite common occurrence, for instance
in either Schwarzschild or Reissner-Nordstr\"om geometries, or more generally whenever $\rho+p_r=0$,
so it is very easy for one to be mislead by those special cases. In particular, in situations of
vanishing flux $\Xi=0$ one obtains the so-called  ``transparency condition'', $\left[G_{\mu
\nu}\; U^{\mu}\,n^{\nu}\right]^+_-=0$, see \cite{Ishak}. The conservation identity,
Eq.~(\ref{conservation}), then reduces to  the simple relationship $\dot{\sigma}=-2\,(\sigma +{\cal
P}) \dot{a}/a$. But in general the ``transparency condition'' does not hold, and one needs the full
version of the conservation equation as given in Eq.~\eqref{E:conservation2}.

Physically, it is particularly important to realize that merely specifying the two bulk geometries $[g_\pm]_{ab}$ is not enough to fully determine the motion of the throat --- some sort of assumption must be made regarding the internal behaviour of the physical material concentrated on the throat itself, and we now turn to investigating this critically important point.

%##########################################################################
\subsection{Integrability of the surface energy density}
%##########################################################################

When does it make sense to assert the existence of a function $\sigma(a)$? Let us start with the
situation in the absence of external forces (we will rapidly generalize this) where the conservation
equation,
\begin{equation}
\dot{\sigma}=-2\,(\sigma +{\cal P}) \dot{a}/a
\end{equation}
can easily be rearranged to
\begin{equation}
{\dot\sigma\over\sigma +{\cal P}} = -2 {\dot a\over a}.
\end{equation}
Assuming a barotropic equation of state ${\cal P}(\sigma)$ for the matter on the wormhole throat,
this can be integrated to yield
\begin{equation}
\int_{\sigma_0}^\sigma {\d\bar\sigma\over\bar\sigma +{\cal P}(\bar\sigma) } = -2 \int_{a_0}^a
{\d\bar a\over \bar a} = -2 \ln(a/a_0).
\end{equation}
This implies that $a$ can be given as some function $a(\sigma)$ of $\sigma$, and by the inverse
function theorem implies over suitable domains the existence of a function $\sigma(a)$. Now this
barotropic equation of state is a rather strong assumption, albeit one that is very often implicitly
made when dealing with thin-shell wormholes (or thin-shell gravastars \cite{gravastar,gravastar2,MartinMoruno:2011rm}, or other thin-shell objects).
As a first generalization, consider what happens if the surface pressure is generalized to be of the form ${\cal P}(a,
\sigma)$, which is not barotropic. Then the conservation equation can be rearranged to be
\begin{equation}
\sigma'= - {2[\sigma +{\cal P}(a,\sigma)]\over a}.
\end{equation}
This is a first-order (albeit nonlinear and non-autonomous) ordinary differential equation, which at
least locally will have solutions $\sigma(a)$. There is no particular reason to be concerned about
the question of global solutions to this ODE, since in applications one is most typically dealing
with linearization around a static solution.

If we now switch on external forces, that is $\Phi^\pm\neq 0$, then one way of guaranteeing integrability would be to demand that the external forces are of the form $\Xi(a,\sigma)$, since then the conservation equation would read
\begin{equation}
\sigma'= - {2[\sigma +{\cal P}(a,\sigma)]\over a} + \Xi(a,\sigma),
\end{equation}
which is again a first-order albeit nonlinear and non-autonomous ordinary differential equation. But
how general is this $\Xi = \Xi(a,\sigma)$ assumption? There are at least two situations where this
definitely holds:
\begin{itemize}
\item If $\Phi_+(a) = \Phi_-(a) = \Phi(a)$, then $\Xi = - \Phi'(a) \; \sigma$, which is explicitly
of the required form.
\item If $b_+(a)=b_-(a)=b(a)$, then  $\Xi = {1\over2}[\Phi'_+(a)+\Phi'_-(a)]\; \sigma$, which is again
explicitly of the required form.
\end{itemize}
But in general we will need a more complicated set of assumptions to assure integrability, and the consequent
existence of some function $\sigma(a)$.
A model that is always \emph{sufficient} (not necessary) to guarantee integrability is to view the exotic material on the
throat as a two-fluid system, characterized by $\sigma_\pm$ and ${\cal P}_\pm$, with two (possibly
independent) equations of state ${\cal P}_\pm(\sigma_\pm)$. Specifically, take
\begin{eqnarray}
\sigma_\pm &=&-\frac{1}{4\pi}\,(K_\pm)^{\theta}_{\;\theta} \,,
\\
{\cal P}_\pm &=&\frac{1}{8\pi}\left\{ (K_\pm)^{\tau}_{\;\tau}+(K_\pm)^{\theta}_{\;\theta}\right\}
\,.
\end{eqnarray}
In view of the differential identities
\begin{equation}
{\d\over\d\tau}\left\{ a \; e^{\Phi_\pm} \; K^{\theta \;\pm }_{\;\;\theta} \right\} = e^{\Phi_\pm}
\; K^{\tau\;\pm}_{\;\;\tau} \; \dot a,
\end{equation}
each of these two fluids is independently subject to
\begin{equation}
{\d\over\d\tau}\left\{ e^{\Phi_\pm} \; \sigma_\pm  \right\} = -{2e^{\Phi_\pm}\over a} \;
\{\sigma_\pm + {\cal P}_\pm\} \; \dot a.
\end{equation}
which is equivalent to
\begin{equation}
\left\{ e^{\Phi_\pm} \; \sigma_\pm  \right\}' = -{2e^{\Phi_\pm}\over a} \; \{\sigma_\pm + {\cal
P}_\pm\}.
\end{equation}
With two equations of state ${\cal P}_\pm(\sigma_\pm)$ these are two nonlinear first-order ordinary differential equations for
$\sigma_\pm$.  These equations are integrable, implicitly defining functions $\sigma_\pm(a)$, at least locally.
Once this is done we define
\begin{equation}
\sigma(a) = \sigma_+(a) + \sigma_-(a),
\end{equation}
and
\begin{equation}
m_s(a) = 4\pi \sigma(a) \; a^2.
\end{equation}
While the argument is more complicated than one might have expected, the end result is easy to
interpret: We can simply \emph{choose} $\sigma(a)$, or equivalently $m_s(a)$, as an arbitrarily
specifiable function that encodes the (otherwise unknown) physics of the specific form of exotic
matter residing on the wormhole throat.

%##########################################################################
\subsection{Equation of motion}
%##########################################################################

To qualitatively analyze the stability of the wormhole, assuming integrability of the surface energy
density, (that is, the existence of a function $\sigma(a)$), it is useful to rearrange
Eq.~(\ref{gen-surfenergy2}) into the form
\begin{eqnarray}
\sqrt{1-\frac{b_{+}(a)}{a}+\dot{a}^{2}}= -\sqrt{1-\frac{b_{-}(a)}{a}+\dot{a}^{2}}
- 4\pi a\,\sigma(a)   \,.\label{def_potential}
\end{eqnarray}
Squaring this equation, rearranging terms to isolate the square root, and finally by squaring once
again, we deduce the thin-shell equation of motion given by
\begin{equation}
{1\over2} \dot{a}^2+V(a)=0  \,,
\end{equation}
where the potential $V(a)$ is given by
\begin{equation}
V(a)= {1\over2}\left\{ 1-{\bar b(a)\over a} -\left[\frac{m_{s}(a)}{2a}\right]^2-\left[\frac{\Delta(a)}{m_{s}
(a)}\right]^2\right\}\,.
   \label{potential}
\end{equation}
Here $m_s(a)=4\pi a^2\,\sigma(a)$ is the mass of the thin shell. The quantities $\bar b(a)$ and
$\Delta(a)$ are defined, for simplicity, as
\begin{eqnarray}
\bar b(a)&=&\frac{b_{+}(a)+b_{-}(a)}{2},\\
\Delta(a)&=&\frac{b_{+}(a)-b_{-}(a)}{2},
\end{eqnarray}
respectively. This gives the potential $V(a)$ as a function of the surface mass $m_s(a)$.
By differentiating with respect to $a$, (using \eqref{E:trick}), we see that  the equation of motion implies
\begin{equation}
\ddot a = -V'(a).
\end{equation}
It is sometimes useful to reverse the logic flow and determine the surface mass as a function of the
potential. Following the techniques used in~\cite{gravastar}, suitably modified for the present
wormhole context, a brief calculation yields
\begin{equation}
m_s^2(a) = 2a^2
\left\{
1 -{\bar b(a)\over a} - 2V(a)
+  \sqrt{ 1- {2\bar b(a)\over a} - 4V(a) +\left[ 2V(a) + {\bar b(a)\over a}\right]^2 - {\Delta(a)^2\over a^2} }
\right\},
\end{equation}
where one is forced to take the positive root to guarantee reality of the surface mass.  Noting that
the radical factorizes we see
\begin{equation}
m_s^2(a) = 2a^2
\left[
1 -{\bar b(a)\over a} - 2V(a)
+  \sqrt{ 1- {b_+(a)\over a} - 2V(a)}\sqrt{ 1- {b_-(a)\over a} - 2 V(a)}
\right],
\end{equation}
and in fact
\begin{equation}
m_s(a) = -a
\left[
  \sqrt{ 1- {b_+(a)\over a} - 2V(a)} +\sqrt{ 1- {b_-(a)\over a} - 2V(a)}
\right],
\end{equation}
with the negative root now being necessary for compatibility with the Lanczos equations. Note the
logic here --- assuming integrability of the surface energy density, if we want a specific $V(a)$
this tells us how much surface mass we need to put on the wormhole throat (as a function of $a$),
which is implicitly making demands on the equation of state of the exotic matter residing on the
wormhole throat. In a completely analogous manner, the assumption of integrability of $\sigma(a)$ implies that
after imposing the equation of motion for the shell one has
\begin{equation}
\sigma(a)=-\frac{1}{4\pi a}\left[\sqrt{1-\frac{b_{+}(a)}{a} - 2 V(a)}+\sqrt{1-\frac{b_{-}(a)}{a}- 2 V(a)}\right],
\label{gen-surfenergy2-onshell}
\end{equation}
while
\begin{eqnarray}
{\cal P}(a)&=&\frac{1}{8\pi a}\left[
\frac{1-2V(a)-aV'(a)-\frac{b_+(a)+ab'_+(a)}{2a}}{\sqrt{1-\frac{b_{+}(a)}{a}-2V(a)}}
+
\sqrt{1-\frac{b_{+}(a)}{a}-2V(a)} \; a\Phi'_{+}(a)
\right. \nonumber\\
&&
\qquad
\left. +
\frac{1-2V(a)-aV'(a)-\frac{b_-(a)+ab'_-(a)}{2a}}{\sqrt{1-\frac{b_{-}(a)}{a}-2V(a)}}
+
\sqrt{1-\frac{b_{-}(a)}{a}-2V(a)} \; a\Phi'_{-}(a)
\right].\label{gen-surfpressure2-onshell}
\end{eqnarray}
and
\begin{equation}
\Xi(a)
=\frac{1}{4\pi a}\, \left[\Phi'_+(a)\sqrt{1-\frac{b_+(a)}{a}-2V(a)} + \Phi'_-(a)\sqrt{1-\frac{b_-(a)}{a}-2V(a)}\right]\,.
\end{equation}
The three quantities $\{\sigma(a),\,{\cal P}(a),\,\Xi(a)\}$ (or equivalently $\{m_s(a),\,{\cal P}(a),\,\Xi(a)\}$) are related by the differential conservation law, so at most two of them are functionally independent.

%##########################################################################
\subsection{Linearized equation of motion}
%##########################################################################

Consider a linearization around an assumed static solution (at $a_0$) to the equation of motion
${1\over2}\dot a^2 + V(a)=0$,  and so also a solution of $\ddot a = -V'(a)$.  Generally a Taylor
expansion of $V(a)$ around $a_0$ to second order yields
\begin{equation}
V(a)=V(a_0)+V'(a_0)(a-a_0)+\frac{1}{2}V''(a_0)(a-a_0)^2+O[(a-a_0)^3]
\,.   \label{linear-potential0}
\end{equation}
But since we are expanding around a static solution, $\dot a_0=\ddot a_0 = 0$, we automatically have
$V(a_0)=V'(a_0)=0$, so it is sufficient to consider
\begin{equation}
V(a)= \frac{1}{2}V''(a_0)(a-a_0)^2+O[(a-a_0)^3]
\,.   \label{linear-potential}
\end{equation}
The assumed static solution at $a_0$ is stable if and only if $V(a)$ has a local minimum at $a_0$,
which requires $V''(a_{0})>0$. This will be our primary criterion for wormhole stability, though it
will be useful to rephrase it in terms of more basic quantities.

For instance, it is extremely useful to express $m_s'(a)$ and $m_s''(a)$ by the following
expressions:
\begin{equation}
m_s'(a) = + {m_s(a)\over a} + {a\over2} \left\{
{ (b_+(a)/a)'+2V'(a)\over\sqrt{1-b_+(a)/a-2V(a)}} +
{(b_-(a)/a)'+2V'(a)\over\sqrt{1-b_-(a)/a-2V(a)}} \right\},
\end{equation}
and
\begin{eqnarray}
m_s''(a) &=& \left\{
{ (b_+(a)/a)'+2V'(a)\over\sqrt{1-b_+(a)/a-2V(a)}} +
{(b_-(a)/a)'+2V'(a)\over\sqrt{1-b_-(a)/a-2V(a)}} \right\}
\nonumber\\
&&
+{a\over4}
\left\{
{ [(b_+(a)/a)'+2V'(a)]^2\over[1-b_+(a)/a-2V(a)]^{3/2}} +
{[(b_-(a)/a)'+2V'(a)]^2\over[1-b_-(a)/a-2V(a)]^{3/2}} \right\}
\nonumber\\
&&
+{a\over2}
\left\{
{ (b_+(a)/a)''+2V''(a)\over\sqrt{1-b_+(a)/a-2V(a)}} +
{(b_-(a)/a)''+2V''(a)\over\sqrt{1-b_-(a)/a-2V(a)}} \right\}.
\end{eqnarray}
Doing so allows us to easily study linearized stability, and to develop a simple inequality on
$m_s''(a_0)$ by using the constraint $V''(a_0)>0$. Similar formulae hold for $\sigma'(a)$, $\sigma''(a)$, for ${\cal P}'(a)$, ${\cal P}''(a)$, and for $\Xi'(a)$, $\Xi''(a)$. In view of the redundancies coming from the relations $m_s(a) = 4\pi\sigma(a) a^2$ and the differential conservation law, the only interesting quantities are  $\Xi'(a)$, $\Xi''(a)$.

For practical calculations, it is extremely useful to  consider the dimensionless quantity
\begin{equation}
{m_s(a)\over a} =4\pi \sigma(a) a = - \left[\sqrt{1-\frac{b_{+}(a)}{a} - 2 V(a)}+\sqrt{1-\frac{b_{-}(a)}{a}- 2 V(a)}\right],
\end{equation}
and then to express $[m_s(a)/a]'$ and $[m_s(a)/a]''$ by the following
expressions:
\begin{equation}
[m_s(a)/a]' = + {1\over2} \left\{
{ (b_+(a)/a)'+2V'(a)\over\sqrt{1-b_+(a)/a-2V(a)}} +
{(b_-(a)/a)'+2V'(a)\over\sqrt{1-b_-(a)/a-2V(a)}} \right\},
\end{equation}
and
\begin{eqnarray}
[m_s(a)/a]'' &=& +{1\over4}
\left\{
{ [(b_+(a)/a)'+2V'(a)]^2\over[1-b_+(a)/a-2V(a)]^{3/2}} +
{[(b_-(a)/a)'+2V'(a)]^2\over[1-b_-(a)/a-2V(a)]^{3/2}} \right\}
\nonumber\\
&&
+{1\over2}
\left\{
{ (b_+(a)/a)''+2V''(a)\over\sqrt{1-b_+(a)/a-2V(a)}} +
{(b_-(a)/a)''+2V''(a)\over\sqrt{1-b_-(a)/a-2V(a)}} \right\}.
\end{eqnarray}
It is similarly useful to consider
\begin{equation}
4\pi \Xi(a) a = \left[\Phi'_+(a)\sqrt{1-\frac{b_+(a)}{a}-2V(a)} + \Phi'_-(a)\sqrt{1-\frac{b_-(a)}{a}-2V(a)}\right]\,.
\end{equation}
for which an easy computation yields:
\begin{eqnarray}
[4\pi\,\Xi(a)\,a]' &=&
+ \left\{
\Phi_+''(a) \sqrt{1-b_+(a)/a-2V(a)} +
\Phi_-''(a) \sqrt{1-b_-(a)/a-2V(a)} \right\}
\nonumber
\\
&&
- {1\over2} \left\{
\Phi_+'(a) { (b_+(a)/a)'+2V'(a)\over\sqrt{1-b_+(a)/a-2V(a)}} +
\Phi_-'(a){(b_-(a)/a)'+2V'(a)\over\sqrt{1-b_-(a)/a-2V(a)}} \right\},
\end{eqnarray}
and
\begin{eqnarray}
[4\pi\,\Xi(a)\,a]'' &=& \left\{
\Phi_+'''(a) \sqrt{1-b_+(a)/a-2V(a)} +
\Phi_-'''(a) \sqrt{1-b_-(a)/a-2V(a)} \right\}
\nonumber\\
&&
- \left\{
\Phi_+''(a) { (b_+(a)/a)'+2V'(a)\over\sqrt{1-b_+(a)/a-2V(a)}} +
\Phi_-''(a){(b_-(a)/a)'+2V'(a)\over\sqrt{1-b_-(a)/a-2V(a)}} \right\}
\nonumber\\
&&
-{1\over4}
\left\{
\Phi_+'(a) { [(b_+(a)/a)'+2V'(a)]^2\over[1-b_+(a)/a-2V(a)]^{3/2}} +
\Phi_-'(a) {[(b_-(a)/a)'+2V'(a)]^2\over[1-b_-(a)/a-2V(a)]^{3/2}} \right\}
\nonumber\\
&&
-{1\over2}
\left\{
\Phi_+'(a) { (b_+(a)/a)''+2V''(a)\over\sqrt{1-b_+(a)/a-2V(a)}} +
\Phi_-'(a) {(b_-(a)/a)''+2V''(a)\over\sqrt{1-b_-(a)/a-2V(a)}} \right\}.
\end{eqnarray}
We shall evaluate these quantities at the assumed stable solution $a_0$.

%%##########################################################################
\subsection{The master equations}
%%##########################################################################
In view of the above, to have a stable static solution at $a_0$ we must have:
\begin{equation}
m_s(a_0) = -a_0
\left\{
  \sqrt{ 1- {b_+(a_0)\over a_0} } +\sqrt{ 1- {b_-(a_0)\over a_0}}
\right\},
 \label{stable_ms}
\end{equation}
while
\begin{equation}
m_s'(a_0) = {m_s(a_0)\over 2 a_0} -{1\over2} \left\{
{ 1 - b_+'(a_0)\over\sqrt{1-b_+(a_0)/a_0}} +
{ 1-  b_-'(a_0)\over\sqrt{1-b_-(a_0)/a_0}} \right\},
 \label{stable_dms}
\end{equation}
and
\begin{eqnarray}
m_s''(a_0) &\geq&
+{1\over4 a_0^3}
\left\{
{ [b_+(a_0)- a_0 b_+'(a_0)]^2\over[1-b_+(a_0)/a_0]^{3/2}} +
{ [b_-(a_0)- a_0 b_-'(a_0)]^2\over[1-b_-(a_0)/a_0]^{3/2}}
\right\}
\nonumber\\
&&
+{1\over2}
\left\{
{b_+''(a_0)\over\sqrt{1-b_+(a_0)/a_0}} +
{b_-''(a_0)\over\sqrt{1-b_-(a_0)/a_0}} \right\}.\;\;\;
  \label{stable_ddms}
\end{eqnarray}
This last formula in particular translates the stability condition $V''(a_0)\geq0$ into a
rather explicit and not too complicated inequality on $m_s''(a_0)$, one that can in particular
cases be explicitly checked with a minimum of effort.

For practical calculations it is more useful to work with $m_s(a)/a$ in which case
\begin{equation}
\left.[m_s(a)/a]'\right|_{a_0} = + {1\over2} \left.\left\{
{ (b_+(a)/a)'\over\sqrt{1-b_+(a)/a}} +
{(b_-(a)/a)'\over\sqrt{1-b_-(a)/a}} \right\}\right|_{a_0},
\end{equation}
and
\begin{eqnarray}
\left.[m_s(a)/a]''\right|_{a_0} &\geq& +{1\over4}
\left.\left\{
{ [(b_+(a)/a)']^2\over[1-b_+(a)/a]^{3/2}} +
{[(b_-(a)/a)']^2\over[1-b_-(a)/a]^{3/2}} \right\}\right|_{a_0}
\nonumber\\
&&
+{1\over2}
\left.\left\{
{ (b_+(a)/a)''\over\sqrt{1-b_+(a)/a}} +
{(b_-(a)/a)''\over\sqrt{1-b_-(a)/a)}} \right\}\right|_{a_0}.
\label{plotdef:ddms}
\end{eqnarray}
In the absence of external forces this inequality (or the equivalent one for $m_s''(a_0)$ above) is the only stability constraint one requires.  However, once one has external forces ($\Phi_\pm\neq 0$),  there is additional information:
\begin{eqnarray}
\left.[4\pi\,\Xi(a)\,a]'\right|_{a_0} &=&
+ \left.\left\{
\Phi_+''(a) \sqrt{1-b_+(a)/a} +
\Phi_-''(a) \sqrt{1-b_-(a)/a} \right\}\right|_{a_0}
\nonumber
\\
&&
- {1\over2} \left.\left\{
\Phi_+'(a) { (b_+(a)/a)'\over\sqrt{1-b_+(a)/a}} +
\Phi_-'(a){(b_-(a)/a)'\over\sqrt{1-b_-(a)/a}} \right\}\right|_{a_0},
\end{eqnarray}
and (provided $\Phi'_\pm(a_0) \geq 0$)
\begin{eqnarray}
\left.[4\pi\,\Xi(a)\,a]''\right|_{a_0} &\leq& \left.\left\{
\Phi_+'''(a) \sqrt{1-b_+(a)/a} +
\Phi_-'''(a) \sqrt{1-b_-(a)/a} \right\}\right|_{a_0}
\nonumber\\
&&
- \left.\left\{
\Phi_+''(a) { (b_+(a)/a)'\over\sqrt{1-b_+(a)/a}} +
\Phi_-''(a){(b_-(a)/a)'\over\sqrt{1-b_-(a)/a}} \right\}\right|_{a_0}
\nonumber\\
&&
-{1\over4}
\left.\left\{
\Phi_+'(a) { [(b_+(a)/a)']^2\over[1-b_+(a)/a]^{3/2}} +
\Phi_-'(a) {[(b_-(a)/a)']^2\over[1-b_-(a)/a]^{3/2}} \right\}\right|_{a_0}
\nonumber\\
&&
-{1\over2}
\left.\left\{
\Phi_+'(a) { (b_+(a)/a)''\over\sqrt{1-b_+(a)/a}} +
\Phi_-'(a) {(b_-(a)/a)''\over\sqrt{1-b_-(a)/a}} \right\}\right|_{a_0}.
   \label{plotdef:ddXi}
\end{eqnarray}

If $\Phi'_\pm(a_0) \leq 0$, we simply have
\begin{eqnarray}
\left.[4\pi\,\Xi(a)\,a]''\right|_{a_0} &\geq& \left.\left\{
\Phi_+'''(a) \sqrt{1-b_+(a)/a} +
\Phi_-'''(a) \sqrt{1-b_-(a)/a} \right\}\right|_{a_0}
\nonumber\\
&&
- \left.\left\{
\Phi_+''(a) { (b_+(a)/a)'\over\sqrt{1-b_+(a)/a}} +
\Phi_-''(a){(b_-(a)/a)'\over\sqrt{1-b_-(a)/a}} \right\}\right|_{a_0}
\nonumber\\
&&
-{1\over4}
\left.\left\{
\Phi_+'(a) { [(b_+(a)/a)']^2\over[1-b_+(a)/a]^{3/2}} +
\Phi_-'(a) {[(b_-(a)/a)']^2\over[1-b_-(a)/a]^{3/2}} \right\}\right|_{a_0}
\nonumber\\
&&
-{1\over2}
\left.\left\{
\Phi_+'(a) { (b_+(a)/a)''\over\sqrt{1-b_+(a)/a}} +
\Phi_-'(a) {(b_-(a)/a)''\over\sqrt{1-b_-(a)/a}} \right\}\right|_{a_0}.
\end{eqnarray}
Note that these last three equations are entirely vacuous in the absence of external forces, which is why they have not appeared in the literature until now.

%##########################################################################
\section{Applications}\label{secIII}
%##########################################################################

In discussing specific examples one now ``merely'' needs to apply the general formalism 
described above. Several examples are particularly important, some to emphasize the features 
specific to possible asymmetry between the two universes used in traversable wormhole 
construction, some to emphasize the importance of NEC nonviolation in the bulk, and some to 
assess the simplifications due to symmetry between the two asymptotic regions.

%##########################################################################
\subsection{Borderline NEC non-violation in the bulk: $\Phi_\pm=0$}
%##########################################################################

On extremely general grounds we know the NEC must be violated somewhere in the spacetime of a 
traversable wormhole. By explicit computation we have seen that NEC violation in the bulk 
region external to the wormhole throat is equivalent to $\Phi'_\pm(r)< 0$.  Thus to minimize 
NEC violations for thin-shell wormholes we should demand $\Phi'_\pm(r)\geq 0$ in the bulk. The 
boundary of this region corresponds to $\Phi_\pm = \hbox{(constant})$, which by a simple 
linear change of the time coordinates can be recast in the form $\Phi_\pm=0$. Thus the 
situation $\Phi_\pm=0$ is particularly interesting not just because of mathematical 
simplicity, but physically interesting because it corresponds to the physical constraint that 
the bulk regions on either side of the wormhole throat be on the verge of violating the NEC. 
This $\Phi_\pm=0$ condition is satisfied, for instance, when the bulk spacetimes are 
Schwarzschild, Reissner--Nordstrom, de~Sitter, Kottler (Schwarzschild--de~Sitter), or 
Reissner--Nordstrom--de~Sitter, though the bulk spacetimes can be more general than any of 
these. Key features of the $\Phi_\pm=0$ traversable wormholes are that in the bulk
\begin{equation}
\rho(r) = - p_r(r) =\frac{1}{8\pi r^{2}}b',\qquad p_{t}(r)=-\frac{1}{16\pi r} b''\,,
\end{equation}
while on the throat
\begin{eqnarray}
\sigma&=&-\frac{1}{4\pi a}\left[\sqrt{1-\frac{b_{+}(a)}{a}
+\dot{a}^{2}}+\sqrt{1-\frac{b_{-}(a)}{a}+\dot{a}^{2}}\right],
\\
{\cal P}&=&\frac{1}{8\pi a}\left[
\frac{1+\dot{a}^2+a\ddot{a}-\frac{b_+(a)+ab'_+(a)}{2a}}{\sqrt{1-\frac{b_{+}(a)}{a}+\dot{a}^{2}}}
+
\frac{1+\dot{a}^2+a\ddot{a}-\frac{b_-(a)+ab'_-(a)}{2a}}{\sqrt{1-\frac{b_{-}(a)}{a}+\dot{a}^{2}}}
\right],
\end{eqnarray}
with the external forces vanishing ($\Xi=0$).
Stability of an assumed static solution at $a_0$ then devolves into a single inequality (\ref{stable_ddms}) or the equivalent (\ref{plotdef:ddms}) being imposed upon $m_s''(a_0)$.

%##########################################################################
\subsection{Mirror symmetry: $b_{\pm}=b$, $\Phi_\pm=\Phi$}
%##########################################################################

Another situation in which significant simplification arises is when the two bulk regions are identical, so that $b_{\pm}=b$, and $\Phi_\pm=\Phi$. Key features of these symmetric traversable wormholes are that on the throat
\begin{eqnarray}
\sigma&=&-\frac{1}{2\pi a}\; \sqrt{1-\frac{b(a)}{a} +\dot{a}^{2}},
\\
{\cal P}&=&\frac{1}{4\pi a}\left[
\frac{1+\dot{a}^2+a\ddot{a}-\frac{b(a)+ab'(a)}{2a}}{\sqrt{1-\frac{b(a)}{a}+\dot{a}^{2}}}
+
\sqrt{1-\frac{b(a)}{a}+\dot{a}^{2}} \; a\Phi'(a)
\right],
\end{eqnarray}
with the external forces being rather simply determined by
\begin{equation}
\Xi = - \Phi'(a) \; \sigma.
\end{equation}

%##########################################################################
\subsection{Thin-shell Schwarzschild wormhole: $b_{\pm}=2M_{\pm}$ and $\Phi_\pm=0$}
%##########################################################################

\begin{figure}[!htb]
  %\centering
  \includegraphics[width=6.0 in]{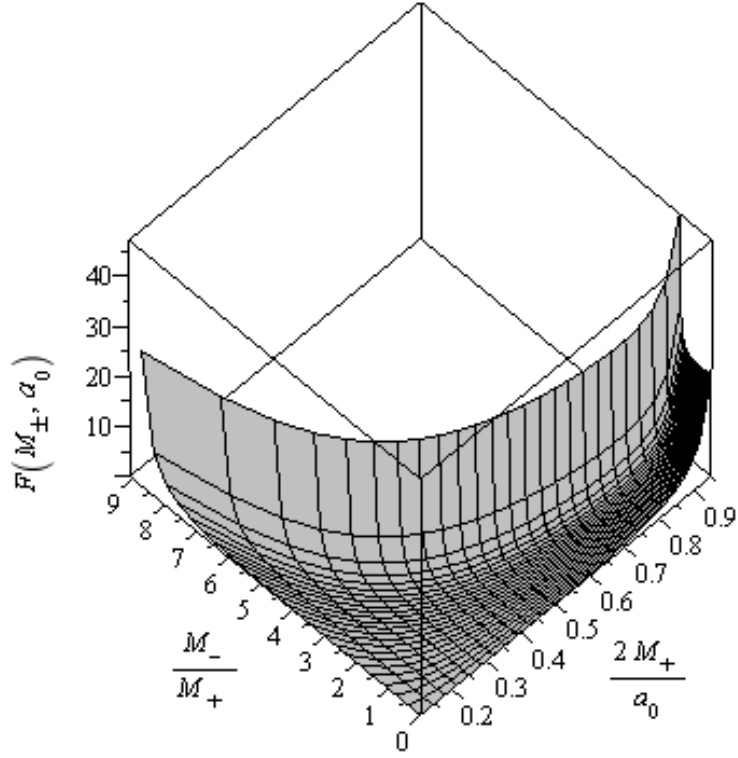}
  \caption{Stability analysis for thin-shell Schwarzschild traversable wormholes.
  The stability region is that above the grey surface depicted in the plot. The grey 
  surface is given by the dimensionless quantity $F(M_\pm,a_0)$, defined by the 
  right-hand-side of inequality (\ref{stability_schw}). We have considered the range 
  $0<x=M_+/a_0 <1$ and $0< y=M_-/M_+ < 1/x$, respectively. Note that a large stability region
  exists for low values of $x=2M_+/a_0$ and of $y=M_-/M_+$. For regions close to the event 
  horizon, $x \rightarrow 1$, the stability region decreases in size and only exists for low values of 
  $y$. See the text for details.}
  \label{ms_schwarz1}
\end{figure}

Consider the situation where both bulk spacetimes on either side of the wormhole throat are portions of Schwarzschild spacetime. Then the metric functions of Eq. (\ref{generalmetric})
are given by
\begin{equation}
b_{\pm}=2M_{\pm},\qquad\qquad \Phi_\pm=0 \,,
\end{equation}
respectively. For this case, inequality (\ref{stable_ddms}) yields the following inequality
\begin{eqnarray}
a_0\,m_s''(a_0) \geq  F(M_\pm,a_0)&=&\left[ \frac{(M_+/a_0)^2}
{\left( 1-2M_+/a_0 \right)^{3/2}}
+ \frac{(M_-/a_0)^2}{\left( 1-2M_-/a_0 \right)^{3/2}} \right] \,.
    \label{stability_schw}
\end{eqnarray}
The dimensionless function $F(M_\pm,a_0)$ is depicted as the grey surface in Fig.
\ref{ms_schwarz1}, and the stability regions are situated above this surface. We have
considered the definition $x=2M_+/a_0$ for convenience, so as to bring infinite $a_0$ in to a finite region of the plot. That is,  $a_0
\rightarrow \infty$ is represented as $x\rightarrow 0$; and $a_0 = 2M_{+}$ is equivalent
$x=1$. Thus, the parameter $x$ is restricted to the range $0<x<1$. We also define the
parameter $y=M_-/M_+$, and from the denominator of inequality (\ref{stability_schw}), we
verify that the parameter $y$ lies within the range $0< y < 1/x$.

From a qualitative analysis of Fig. \ref{ms_schwarz1}, we note that large stability regions exist for small values of $x$ and of $y$. The stability regions decrease for large values of $y$, i.e., for $M_- \gg M_+$, and for large values of $x$, i.e., for regions close to the event horizon. This analysis generalizes previous work, where the case of $M_-=M_+$ was studied \cite{Poisson}. The stability regions decrease in the vicinity of the event horizon, $x \rightarrow 1$, and only exist for low values of $y$.

%##########################################################################
\subsection{Thin-shell Reissner-Nordstr\"{o}m wormholes}
%##########################################################################

The Reissner-Nordstr\"{o}m spacetime is the unique spherically symmetric
solution of the vacuum Einstein-Maxwell coupled equations. Its metric is given
by
\begin{equation}
ds^2=-\left( 1-\frac{2M}{r}+\frac{Q^2}{r^2}\right) dt^2+\left( 1-
\frac{2M}{r}+\frac{Q^2}{r^2}\right) ^{-1}\,dr^2+r^2\left( d\theta
^2+\sin ^2\theta d\phi ^2\right),  \label{RNmetric}
\end{equation}
where $M$ is the mass and $Q^{2}$ is the sum of the squares of the
electric ($Q_{\rm{E}}$) and magnetic ($Q_{\rm{M}}$) charges. In a
local orthonormal  frame, the nonzero components of the
electromagnetic field tensor are
$F^{\hat{t}\hat{r}}=E^{\hat{r}}=Q_{\rm{E}}/r^{2}$ and
$F^{\hat{\theta }\hat{\varphi }}=B^{\hat{r}}=Q_{\rm{M}}/r^{2}$. If
$\left| Q\right| \leq M$, an event horizon is present, at a location
given by
\begin{equation}
r_{b}=M+\sqrt{M^{2}-Q^{2}}.  \label{RNevent-hor}
\end{equation}
If $\left| Q\right| >M$ we have a naked singularity. In the above, we have dropped the $\pm$ subscript as to not overload the notation.

\begin{figure}[!htb]
  \centering
  \includegraphics[width=3.5 in]{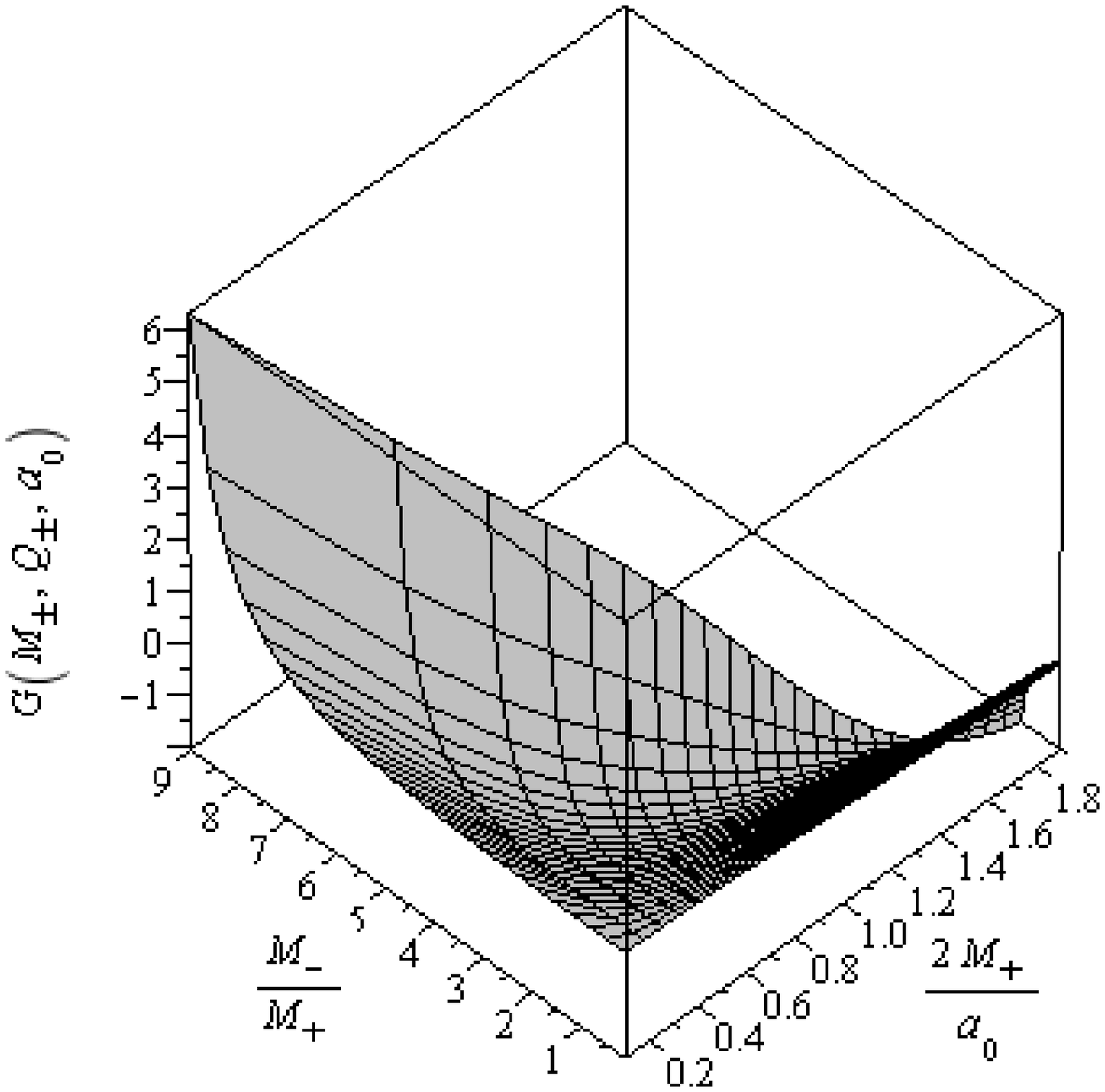}
  \includegraphics[width=3.5 in]{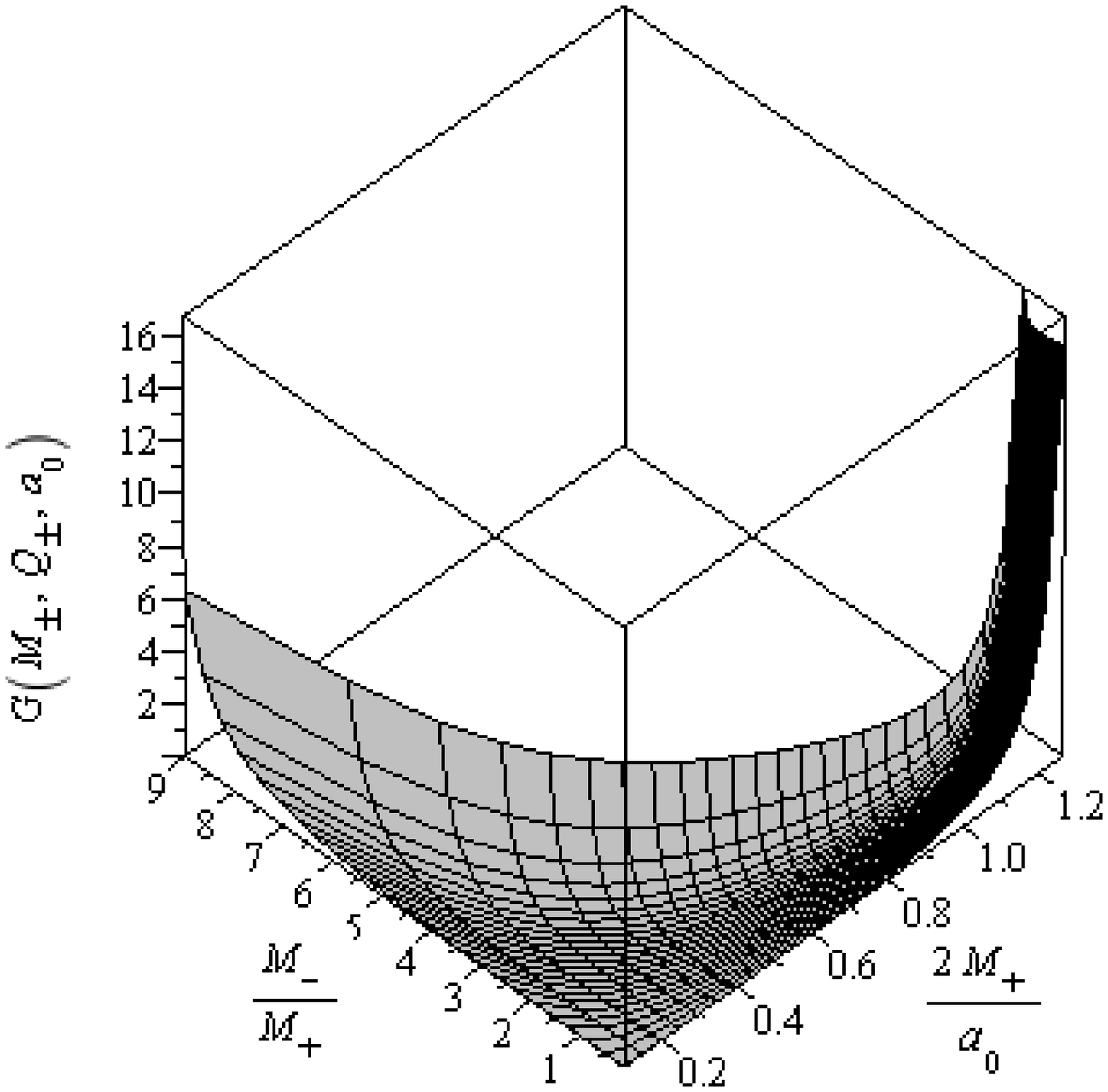}
  \caption{Stability analysis for thin-shell Reissner-Nordstr\"{o}m traversable wormholes.
  The stability region is that above the grey surface. The grey
  surface is given by the dimensionless quantity $G(M_\pm ,Q_\pm ,a_0)$, defined in Eq.
  (\ref{stability_reissner}). The values of $|Q|/M_+=1$ and $|Q|/M_+=\sqrt{3}/2$ are depicted
  in the left and right plots, respectively. Note that for decreasing values of $|Q|/M_+$, the
  stability regions decrease substantially for low values of $M_-/M_+$ and for high values of
  $2M_+/a_0$. See the text for more details.}
  \label{ms_reissner_Q1}
\end{figure}

As in the Schwarzschild solution, one may construct a thin-shell wormhole solution, using the cut-and-paste procedure. For this case, inequality (\ref{stable_ddms}) yields the following dimensionless quantity
\begin{eqnarray}
a_0\,m_s''(a_0) \geq  G(M_\pm ,Q_\pm ,a_0)&=&
\left[
{( M_+^2-Q_+^2)/a_0^2\over \left( 1-2M_+/a_0 +Q_+^2/a_0^2\right)^{3/2}}
+
{( M_-^2-Q_-^2)/a_0^2\over \left( 1-2M_-/a_0 +Q_-^2/a_0^2\right)^{3/2}}
\right]
    \,.
    \label{stability_reissner}
\end{eqnarray}
The surface $G(M_\pm ,Q_\pm ,a_0)$ is shown in Fig. (\ref{ms_reissner_Q1}), and the stability regions, $a_0\,m_s''(a_0)$, are depicted above this surface.

As in the previous example, consider the following definitions: (i) $x=2M_+/a_0$, in order to 
bring in infinity, i.e., $a_0 \rightarrow \infty$ is represented as $x\rightarrow 0$; and (ii) 
$y = M_-/M_{+}$. Using these definitions, and considering for simplicity that $Q_+=Q_-=Q$, we 
have the following ranges
\begin{equation}
0< x < \frac{2}{1+\sqrt{1-Q^2/M_+^2}}\,, \qquad 0< y< \frac{1+x^2(Q^2/4M_+^2)}{x} \,.
\end{equation}
The first range can be deduced from the condition that $r_b<a_0 <+\infty$; the second range can easily be deduced from the denominators of inequality (\ref{stability_reissner}).

To now analyse the stability regions, we simply consider specific values for $|Q|/M_+$. The values of $|Q|/M_+=1$ and $|Q|/M_+=\sqrt{3}/2$ are depicted in the left and right plots, respectively, of Fig. \ref{ms_reissner_Q1}. Note that for decreasing values of $|Q|/M_+$, the stability regions decrease substantially for low values of $M_-/M_+$ and for high values of $M_+/a_0$. More specifically, if one were to construct thin-shell Reissner-Nordstr\"{o}m wormholes with the junction interface close to the event horizon, one would need high values for the charge in order to have stable solutions.

%##########################################################################
\subsection{Thin-shell variant of the Ellis wormhole: $b_{\pm}=R_{\pm}^2/r$}
%##########################################################################

It is perhaps instructive to consider an explicit case which violates some of the energy conditions in the bulk. Consider, for instance, the case given by the following shape functions: $b_{\pm}=R_{\pm}^2/r$, (so that $b_\pm(r)/r = R_\pm^2/r^2$), which were used in the Ellis wormhole \cite{ellis}. (Note that Ellis' terminology is slightly unusual, and we have rephrased his work in more usual terminology.) 

%##########################################################################
\subsubsection{Zero momentum flux: $\Phi_\pm =0$}
%##########################################################################

\begin{figure}[!t]
  %\centering
  \includegraphics[width=5.0 in]{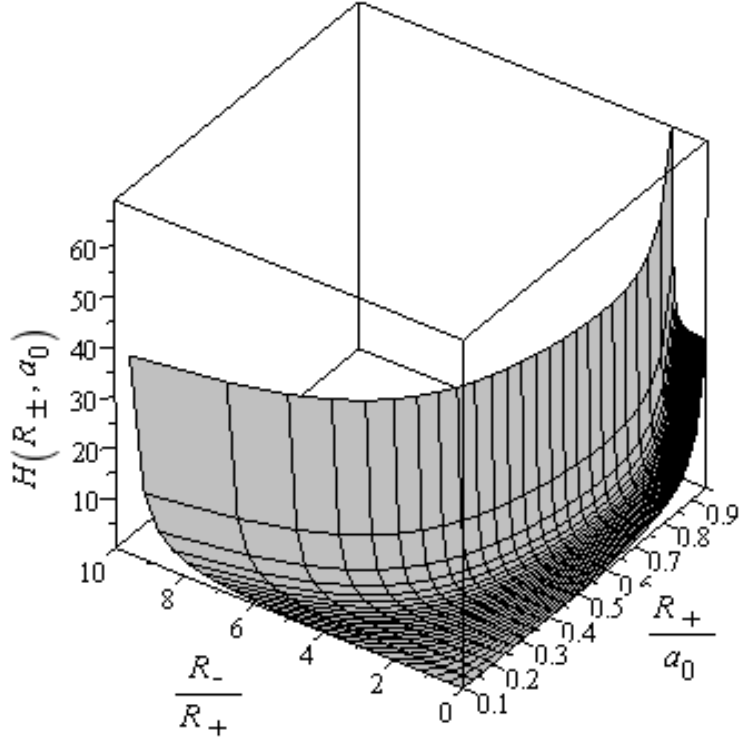}
  \caption{The plot depicts the stability region for the thin-shell variant of the Ellis wormhole, for which $b_{\pm}=R_{\pm}^2/r$. The grey surface is given by the dimensionless quantity 
  $H(R_\pm,a_0)$, given by the right-hand-side of inequality (\ref{stability_ellis}). The 
  stability region, represented by the function $a_0\,m_s''(a_0)$, is that above 
  the grey surfaces depicted in the plot. We have considered the range $0<x=R_+/a_0 <1$ and 
  $0<y=R_-/R_+ < 1/x $. Note that large stability regions exist for small values of $x$ and of 
  $y$. The stability regions decrease for large values of $y$, i.e., for $R_- \gg R_+$, and 
  for large values of $x$. See the text for details.}
  \label{ms_ellis}
\end{figure}

As an initial step, consider the absence of external forces, so that $\Xi=0$. For this case one verifies that the NEC is borderline satisfied in the bulk. However, the WEC is violated as one necessarily has negative energy densities, which is transparent from the following stress-energy profile
\begin{equation}
\rho(r) = - p_r(r) = - p_t(r)=-\frac{R_\pm^2}{8\pi r^{4}}\,.
\end{equation}
Stability regions are dictated by inequality (\ref{stable_ddms}), which yields the following dimensionless quantity
\begin{eqnarray}
a_0\,m_s''(a_0) \geq H(R_\pm,a_0)=
\frac{(R_+/a_0)^2}{\left( 1-R^2_+/a^2_0 \right)^{3/2}}
+
\frac{(R_-/a_0)^2}{\left( 1-R^2_-/a^2_0 \right)^{3/2}}
\,.
\label{stability_ellis}
\end{eqnarray}
The function $H(R_\pm,a_0)$ is depicted as the grey surface in Fig. \ref{ms_ellis} and the stability regions lie above this surface. We consider the definition $x=R_+/a_0$ for convenience, so as to bring in infinity, i.e., $a_0 \rightarrow \infty$ is represented as $x\rightarrow 0$; and $a_0 = R_+$ is equivalent to $x=1$, so that the range for $x$ is given by $0<x<1$. We also consider the definition $y=R_-/R_+$, so that from the denominator of $H(R_\pm,a_0)$, given by Eq. $(\ref{stability_ellis})$, we verify that the range of $y$ is provided by $0<y<1/x$.
The stability analysis is similar to the Schwarzschild thin-shell wormhole case, in that large
stability regions exist for small values of $x$ and of $y$. The stability regions decrease for
large values of $y$, i.e., for $R_- \gg R_+$, and for large values of $x$.

%##########################################################################
\subsubsection{Non-zero external forces: $\Phi_\pm=-R_\pm/r$}
%##########################################################################

In order to generalize the previous specific example, we now consider a case  with external forces. Specifically, let us consider the 
following functions 
\begin{equation}
\Phi_\pm=-\frac{R_\pm}{r}\,.
\end{equation}
These functions imply that $\Phi_\pm ({a_0}) > 0$, so that in addition to the stability 
condition given by inequality (\ref{stable_ddms}), one needs to take into account the 
stability condition dictated by (\ref{plotdef:ddXi}). The latter inequality yields the 
following dimensionless quantity
\begin{eqnarray}
a_0^3\,\left[4\pi a \Xi(a)\right]'' \geq H_2(R_\pm,a_0)&=&
{(6R_+/a_0-19R_+^3/a_0^3+12R_+^5/a_0^5)\over\left( 1-R^2_+/a^2_0 \right)^{3/2}}
+
{(6R_-/a_0-19R_-^3/a_0^3+12R_-^5/a_0^5)\over\left( 1-R^2_-/a^2_0 \right)^{3/2}}\,.
    \label{stability_ellis_Xi}
\end{eqnarray}

Note that as before, inequality (\ref{stable_ddms}) yields the dimensionless quantity given by 
inequality (\ref{stability_ellis}), which is depicted as the grey surface in Fig 
\ref{ms_ellis}. The stability regions are depicted above this surface. We consider once again 
the definition $x=R_+/a_0$, so that the range for $x$ is given by $0<x<1$; and the definition 
$y=R_-/R_+$, so that as before the parameter $y$ lies within the range $0<y<1/x$.

\begin{figure}[!thb]
  %\centering
  \includegraphics[width=3.5 in]{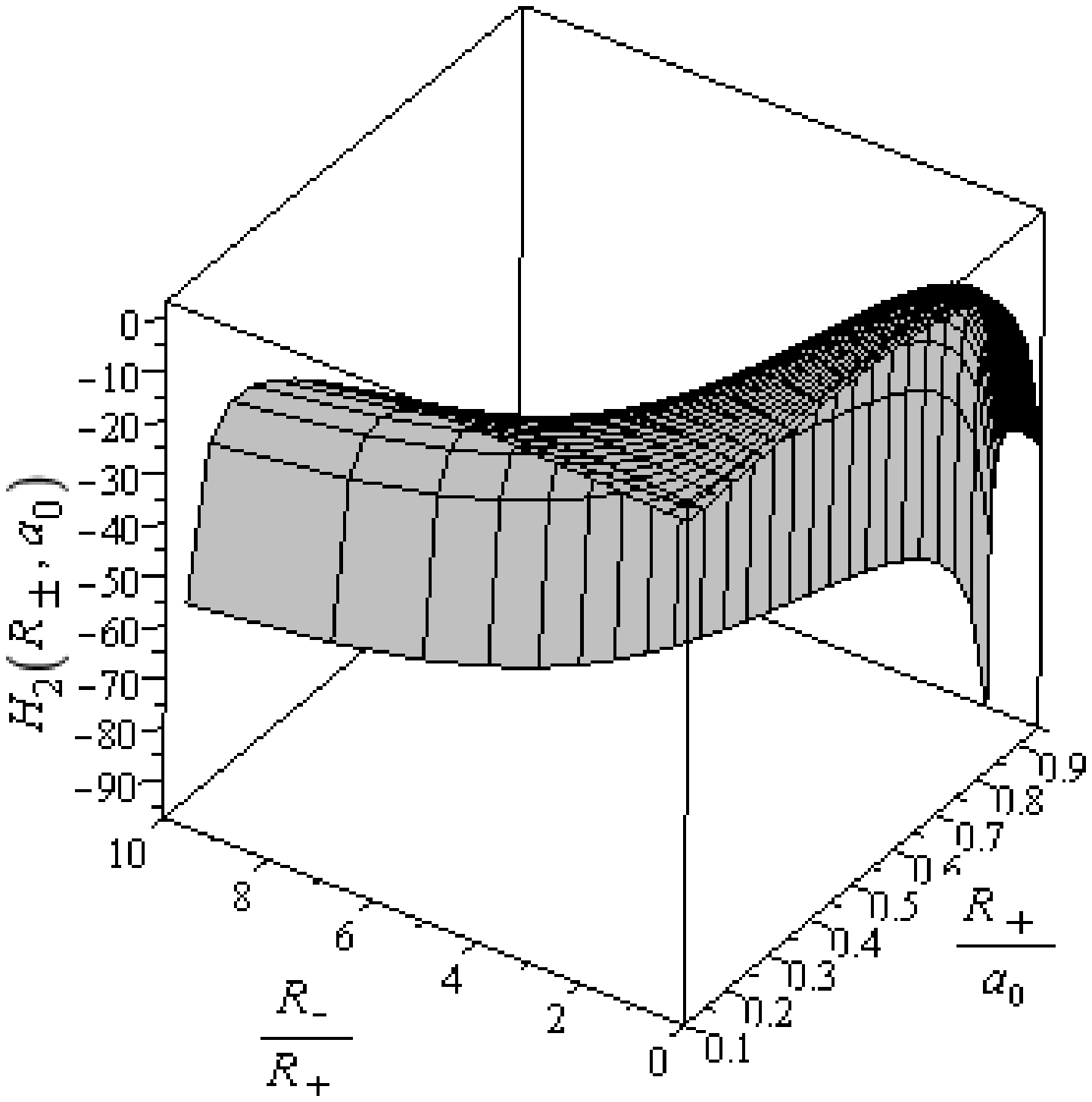}
  \includegraphics[width=3.5 in]{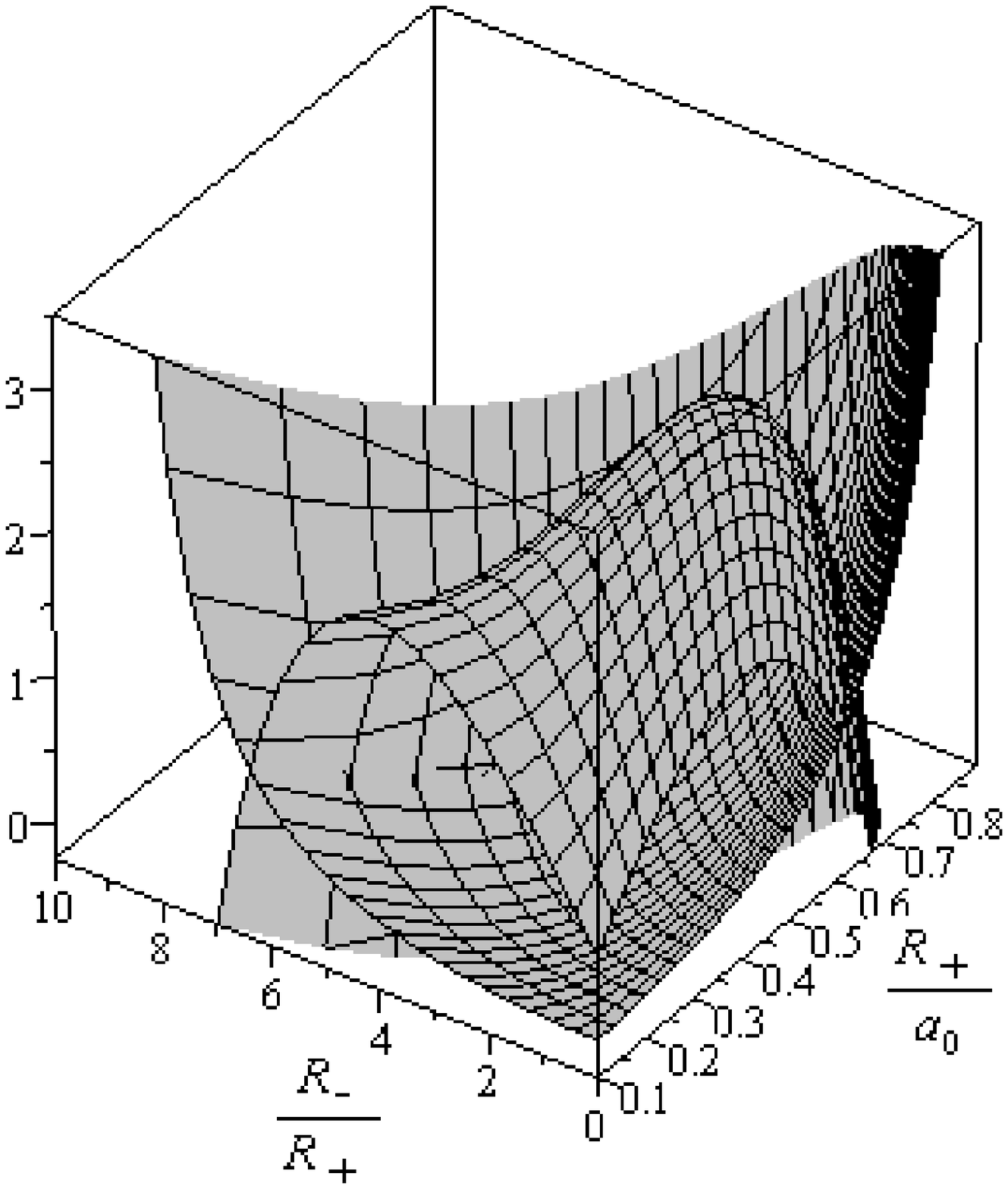}
  \caption{The plot depicts the stability regions for the thin-shell variant of the Ellis 
  wormhole, with  $b_{\pm}=R_{\pm}^2/r$, but now in the presence of non-zero external forces. 
  The latter non-zero momentum flux arises from the functions given by $\Phi_\pm=-R_\pm/r$. 
  The grey surface depicted in the left plot, is given by $H_2(R_\pm,a_0)$, i.e.,
  the right-hand-side of inequality (\ref{stability_ellis_Xi}), and the stability regions are 
  given below this surface. We have considered, as before, the range
  $0<x=R_+/a_0 <1$ and $0< y=R_-/R_+ < 1/x$. Now, collecting the results, note that the 
  stability regions are given from the regions above the surface in Fig. \ref{ms_ellis}, and 
  below the surface provided by the left plot.
  These stability regions are depicted in the right plot in between the grey surfaces, 
  represented by the functions $H(R_\pm,a_0)$ and $H_2(R_\pm,a_0)$, respectively. Note the 
  absence of stability regions for high values of $x$ and $y$. See the text for more details.}
  \label{ms_ellis_Xi}
\end{figure}

Now, in addition to the imposition of inequality (\ref{stability_ellis}), depicted above the 
surface in Fig. \ref{ms_ellis}, the stability regions are also restricted by the condition  
(\ref{stability_ellis_Xi}), depicted in the left plot of Fig. \ref{ms_ellis_Xi}. In the latter, 
the stability regions are given below the grey surface. Collecting the results outlined above, 
note that the stability regions are given from the regions above the surface in Fig. 
\ref{ms_ellis}, and below the surface provided by the left plot of Fig. \ref{ms_ellis_Xi}.
These stability regions are depicted in the right plot of Fig. \ref{ms_ellis_Xi} in between 
the grey surfaces, represented by the functions $H(R_\pm,a_0)$ and $H_2(R_\pm,a_0)$, 
respectively. Note the absence of stability regions for high values of $x$ and $y$.

%###########################################################
\subsection{New toy model: $b_{\pm}=\sqrt{r R_{\pm}}$}
%###########################################################

%##########################################################################
\subsubsection{Zero momentum flux: $\Phi_\pm =0$}
%##########################################################################

Another new and interesting toy model is given by the following 
\begin{equation}
b_{\pm}=\sqrt{r R_{\pm}} \qquad {\rm and} \qquad \Phi_\pm =0\,, \qquad \hbox{so that} \qquad b_\pm(r)/r = \sqrt{R_\pm/r}\,.
\end{equation}
The bulk stress-energy profile is given by
\begin{equation}
\rho(r) = - p_r(r) = 4p_t(r)=\frac{1}{16\pi r^2}\sqrt{\frac{R_\pm}{r}}\,.
\end{equation}
Note that the NEC and WEC are satisfied throughout the bulk.
For this case, inequality (\ref{stable_ddms}) yields the following dimensionless quantity
\begin{eqnarray}
 a_{0} m_{s}''(a_0)\geq  I(R_{\pm},a_{0})=
\frac{ 3R_+/a_0 -2 \sqrt{R_+/a_0} }{16\left(1- \sqrt{R_+/a_0} \right)^{3/2}}
+
\frac{ 3R_-/a_0 -2 \sqrt{R_-/a_0} }{16\left(1- \sqrt{R_-/a_0} \right)^{3/2}}\,.
 \label{new-case1}
\end{eqnarray}

\begin{figure}[!htb]
  %\centering
  \includegraphics[width=5.0 in]{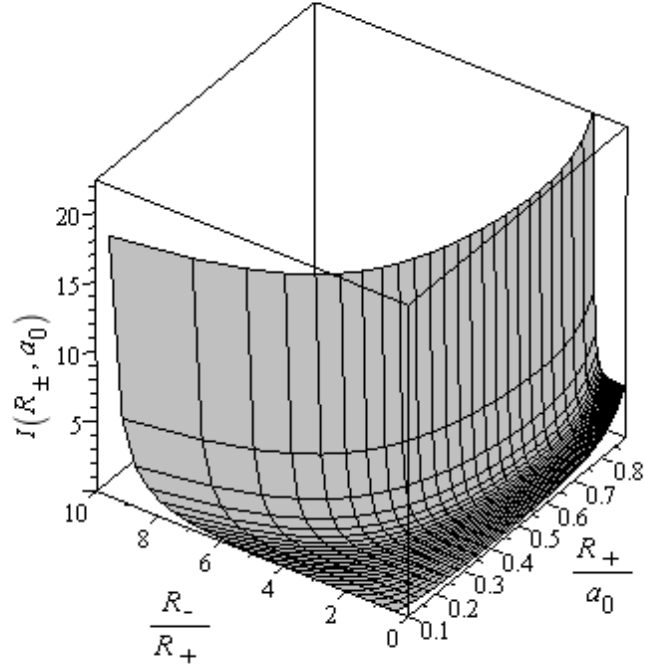}
  \caption{The stability regions for the toy-model thin-shell wormhole with $b_{\pm}=\sqrt{r R_{\pm}}$,  $\Phi_\pm =0$ are given above the grey surfaces depicted in the plot. The 
  grey surface is given by the dimensionless quantity $I(R_{\pm},a_{0})$, given by the 
  right-hand-side of
  inequality (\ref{new-case1}). We have considered the range $0<x=R_+/a_0 <1$ and $0< 
  y=R_-/R_+ < 1/x $. Large stability regions exist for small values of $x$ and of $y$. Note 
  that the stability regions decrease for large values of $y$, i.e., for $R_- \gg R_+$, and 
  for large values of $x$.}
  \label{new-case}
\end{figure}

The stability regions are presented in Fig. \ref{new-case}. As in the previous examples, we 
consider the definition $x=R_+/a_0$ for convenience, so that the range for $x$ is given by 
$0<x<1$. We also consider the definition $y=R_-/R_+$, so that the range of $y$ is provided by 
$0<y<1/x$. Note that, as in the previous example of the thin-shell variant of the Ellis 
wormhole, large stability regions exist for small values of $x$ and of $y$. The stability 
regions decrease for large values of $y$, i.e., for $R_- \gg R_+$, and for large values of 
$x$.

%##########################################################################
\subsubsection{Non-zero external forces: $\Phi_\pm=-R_\pm/r$}
%##########################################################################

As in the previous application, it is useful to generalize the case of zero external forces. In order to have the presence of a non-zero momentum flux term, consider once again the following functions 
\begin{equation}
\Phi_\pm=-\frac{R_\pm}{r}\,,
\end{equation}
which imply that $\Phi_\pm ({a_0}) > 0$. Thus, in addition to the stability 
condition given by inequality (\ref{stable_ddms}), one needs to take into account the 
stability condition dictated by (\ref{plotdef:ddXi}). From the latter inequality, one deduces 
the following dimensionless quantity
\begin{eqnarray}
a_0^3\,\left[4\pi a \Xi(a)\right]'' \geq I_2(R_\pm,a_0)&=&
{(96R_+/a_0-214(R_+/a_0)^{3/2}+117R_+^2/a_0^2)\over {16\left( 1-\sqrt{R_+/a_0} \right)^{3/2}}}
  \nonumber  \\
&&+
{(96R_-/a_0-214(R_-/a_0)^{3/2}+117R_-^2/a_0^2)\over {16\left( 1-\sqrt{R_-/a_0} \right)^{3/2}}}\,.
    \label{stability_toy_Xi}
\end{eqnarray}

\begin{figure}[!htb]
  %\centering
  \includegraphics[width=3.5 in]{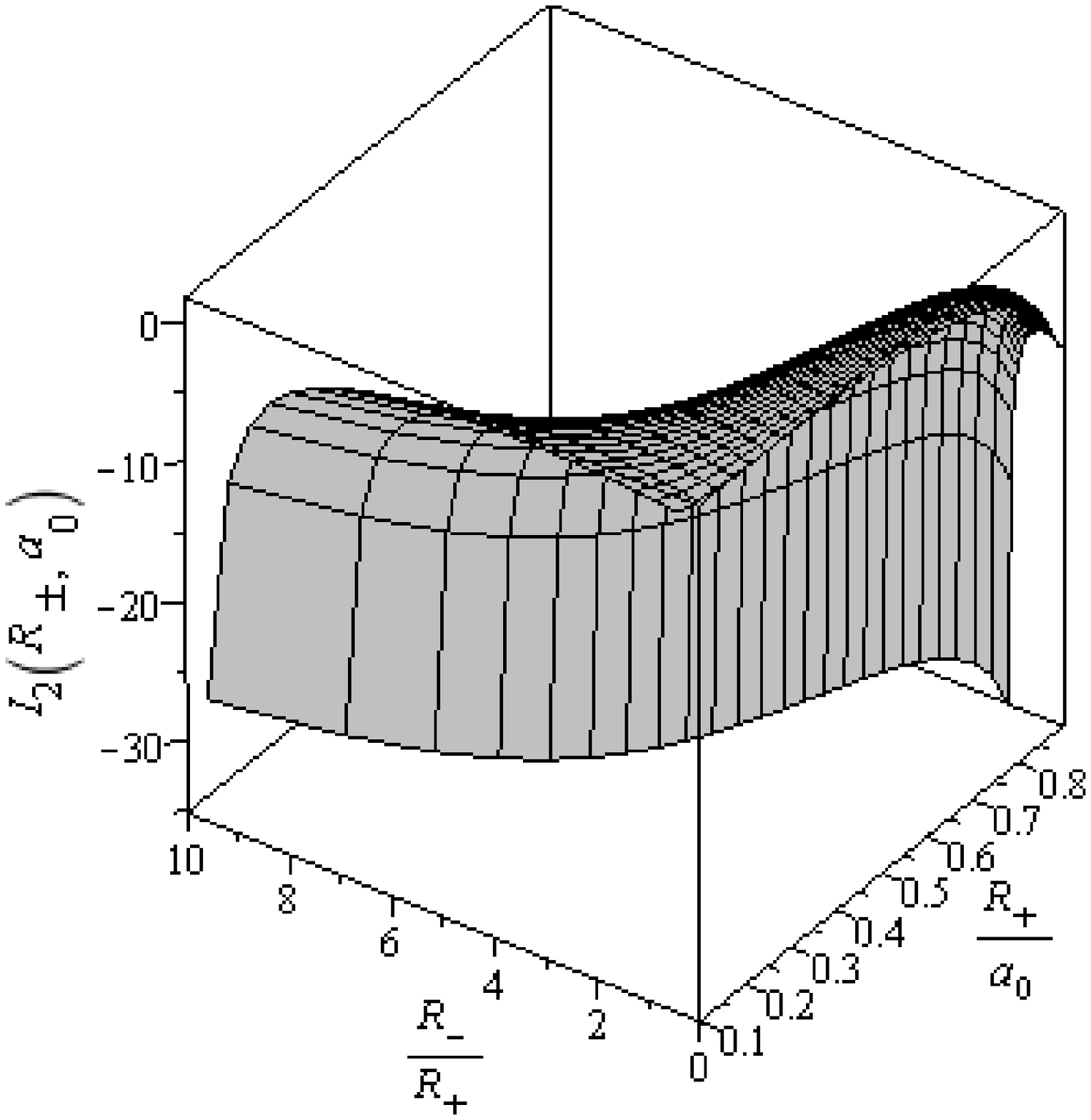}
  \includegraphics[width=3.5 in]{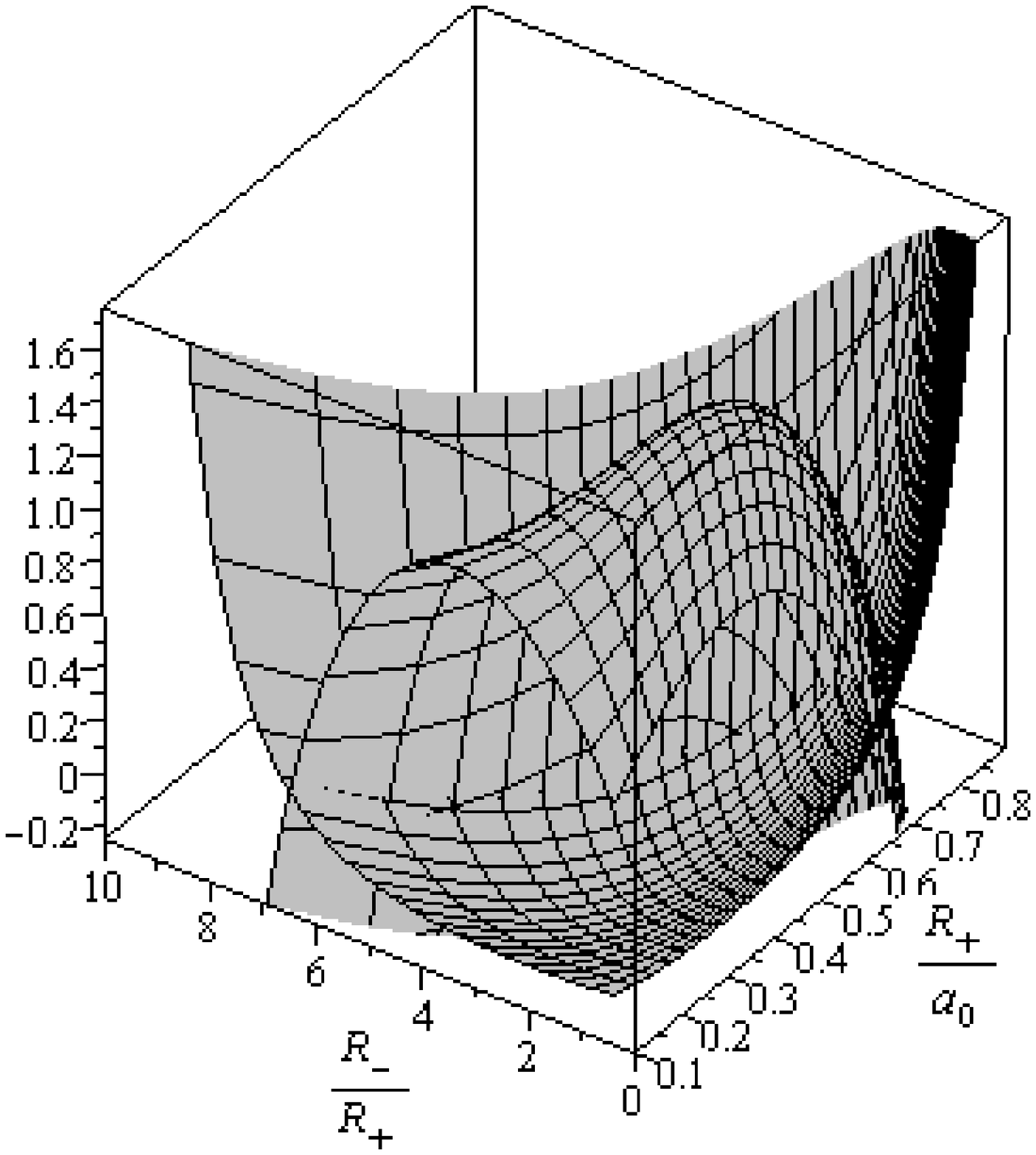}
  \caption{The plot depicts the stability regions for the toy model thin-shell, with  
  $b_{\pm}=\sqrt{R_{\pm}r}$, but now in the presence of non-zero external forces. 
  The latter 
  non-zero momentum flux arises from the functions given by $\Phi_\pm=-R_\pm/r$. The grey 
  surface depicted in the left plot, is given by $I_2(R_\pm,a_0)$, i.e.,
  the right-hand-side of inequality (\ref{stability_toy_Xi}), and the stability regions are 
  given below this surface. We have considered, as before, the range
  $0<x=R_+/a_0 <1$ and $0< y=R_-/R_+ < 1/x$. Now, collecting the results, note that the 
  stability regions are given from the regions above the surface in Fig. \ref{new-case}, and 
  below the surface provided by the left plot.
  These stability regions are depicted in the right plot in between the grey surfaces, 
  represented by the functions $I(R_\pm,a_0)$ and $I_2(R_\pm,a_0)$, respectively. Note the 
  absence of stability regions for high values of $x$ and $y$. See the text for more details.}
  \label{ms_toy_Xi}
\end{figure}

The stability regions are dictated by inequalities (\ref{new-case1}) and 
(\ref{stability_toy_Xi}). These are depicted above the grey surfaces in Fig. \ref{new-case} 
and below the surface of the left plot in Fig. \ref{ms_toy_Xi}. The final stability regions 
are depicted in the right plot of Fig. \ref{ms_toy_Xi} in between the grey surfaces, 
represented by the functions $H(R_\pm,a_0)$ and $H_2(R_\pm,a_0)$, respectively. As in the 
thin-shell variant of the Ellis wormhole, note the absence of stability regions for high 
values of $x$ and $y$.

%##########################################################################
\subsection{Thin-shell charged dilatonic wormhole}
%##########################################################################

Consider a combined gravitational-electromagnetism-dilaton system \cite{dirtyBH, dilaton1, dilaton2}, described by the following Lagrangian
\begin{equation}
{\cal L}=\sqrt{-g}\left\{-R/8\pi + 2 (\nabla\psi)^2 + e^{2\psi} F^2/4\pi \right\}.
\end{equation}
In Schwarzschild coordinates the solution corresponding to an
electric monopole is given by the following line element \cite{dirtyBH, dilaton1, dilaton2},
\begin{eqnarray}
ds^2 =- \left( 1 - {2M\over \beta +\sqrt{r^2+\beta^2} } \right)\, dt^2
         + \left( 1 - {2M\over \beta +\sqrt{r^2+\beta^2} } \right)^{-1}
          {r^2\over r^2+\beta^2}\;dr^2
	 + r^2(d\theta^2 + \sin^2\theta \; d\varphi^2)\,,
\end{eqnarray}
where we have dropped the subscripts $\pm$ for notational convenience.
The non-zero component of the electromagnetic tensor is given by $F_{\hat{t} \hat{r}} = Q/r^2$, and the dilaton field is given by $e^{2\psi} = 1 - {Q^2/ M(\beta+\sqrt{r^2+\beta^2})}$.
The parameter $\beta$ is defined by $\beta\equiv Q^2/2M$. 

In terms of the formalism developed in this paper, the metric functions are provided by
\begin{eqnarray}
b(r) &&=r \left[1-\left( 1 + {\beta^2\over r^2} \right)
                 \left( 1 - {2M\over \beta +\sqrt{r^2+\beta^2} } \right)\right], \\
\Phi(r) &&=-\frac{1}{2}\ln \left( 1 + {\beta^2\over r^2} \right).
\end{eqnarray}
An event horizon exists at $r_b = 2M\sqrt{1-\beta/M}$.
Note that $\Phi'(r)$ is given by
\begin{equation}
\Phi'(r)=\frac{\beta^2}{a(\beta^2+a^2)}\,,
\end{equation}
which is positive throughout the spacetime. Thus, the stability regions are restricted by the 
inequalities (\ref{stable_ddms}) and (\ref{plotdef:ddXi}).

We consider for simplicity $\beta_{+}=\beta_{-}$. The expressions for inequalities 
(\ref{stable_ddms}) and (\ref{plotdef:ddXi}) are extremely lengthy, so we will not write them 
down explicitly.
We define the following parameters $x=2M_+/a$ and $y=M_-/M_+$. The junction interface lies 
within the range $r_b<a<\infty$, so that the range of the parameter $x$ is given by
\begin{equation}
0<x<\frac{1}{2\sqrt{1-\frac{Q^2}{2M_+^2}}} \,.
\end{equation}
In the following analysis, we only consider the regions within the range $0<y \leq 1$, as the 
stability regions lie within this range, as will be shown below. 

\begin{figure}[!htb]
  %\centering
  \includegraphics[width=3.5 in]{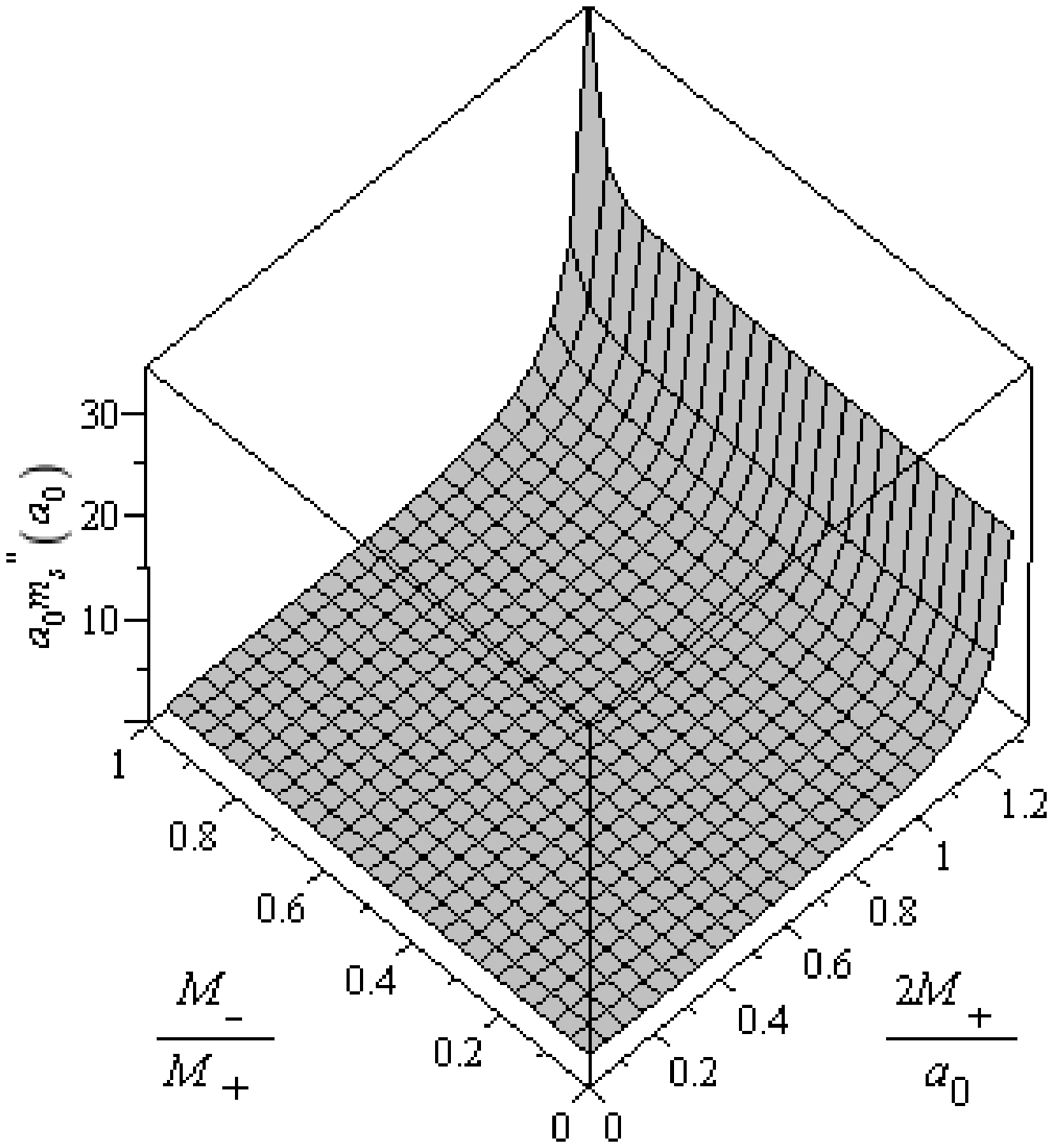}
  \includegraphics[width=3.5 in]{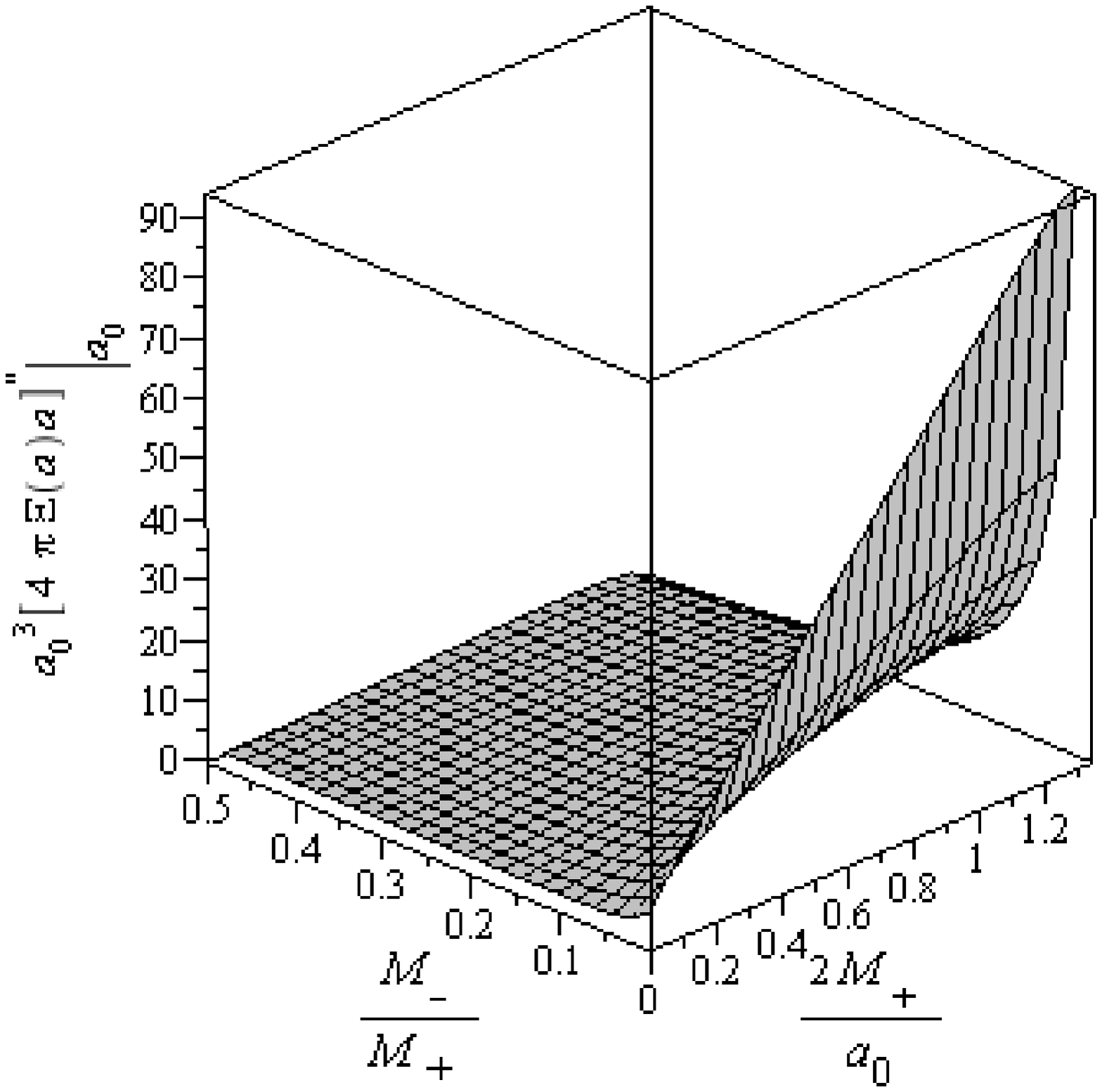}
  \caption{Thin shell dilaton wormhole for $\beta=1/2$. The left plot describes the stability 
  regions above the grey surface, given by (\ref{stable_ddms}), while the right plot describes 
  the stability regions below the grey surface, i.e., inequality (\ref{plotdef:ddXi}). From 
  the left plot, one verifies shows that the stability regions 
  decrease significantly for increasing values of $x$, within the considered range of $0<y<1$. 
  In counterpart, the right plot shows that the stability regions decrease significantly for 
  decreasing values of $y$. These latter stability regions further decrease for increasing 
  values of $x$. See the text for more details.}
  \label{dilaton-blackhole1}
\end{figure}

\begin{figure}[!htb]
  %\centering
  \includegraphics[width=3.5 in]{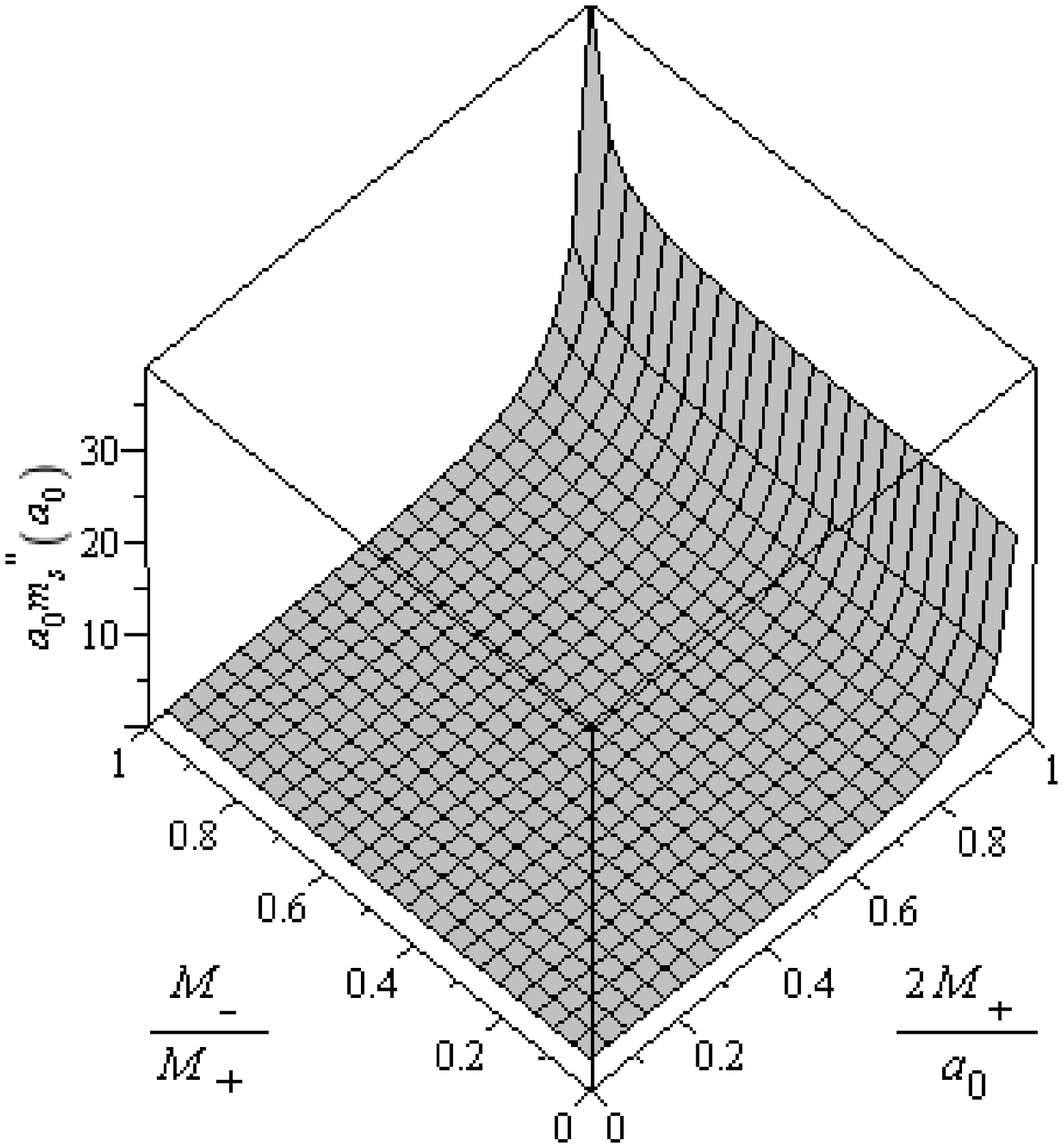}
  \includegraphics[width=3.5 in]{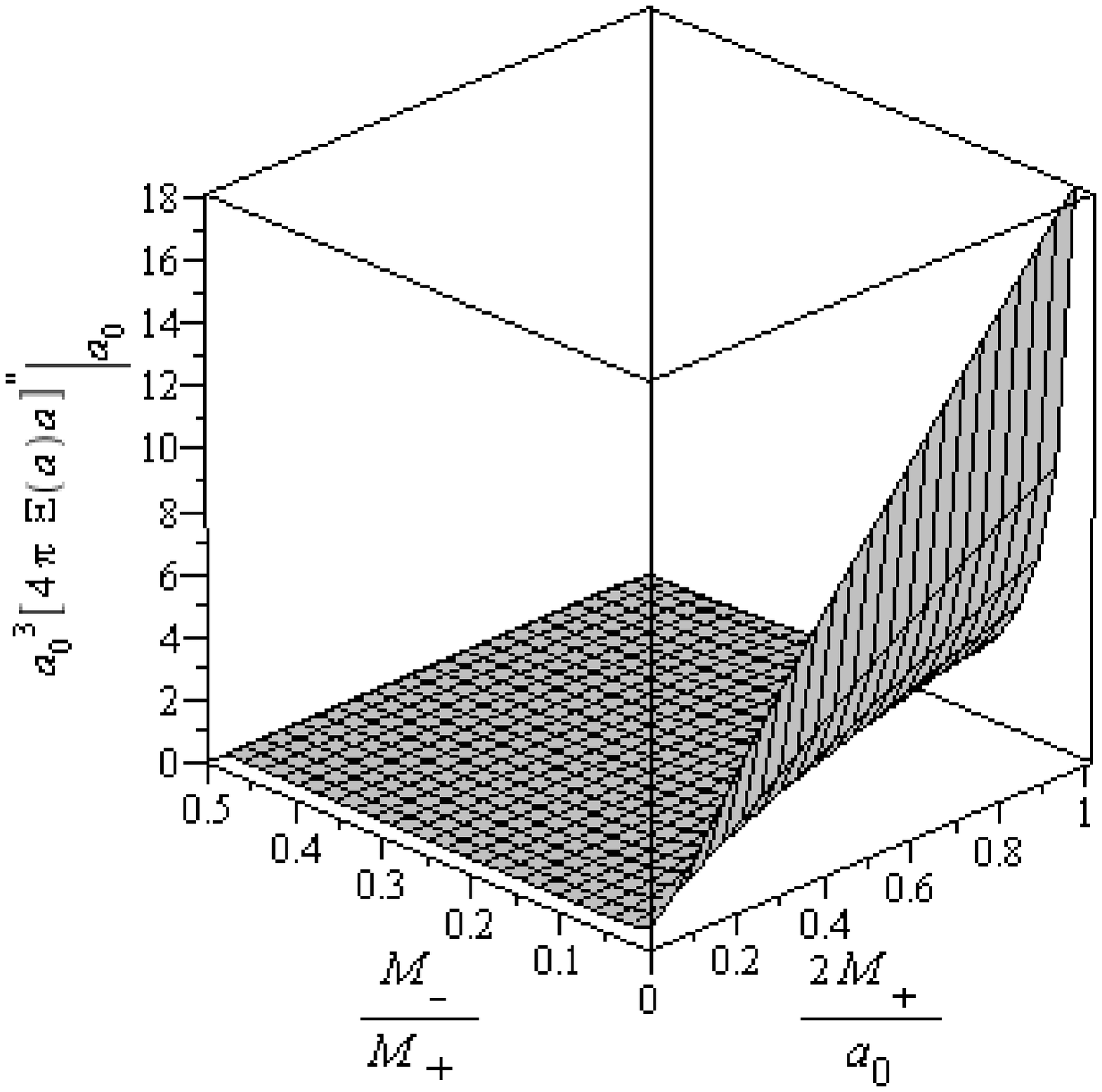}
  \caption{Thin shell dilaton wormhole for $\beta=1/8$. The left plot describes the stability 
  regions above the grey surface, given by (\ref{plotdef:ddms}), while the right plot 
  describes the stability regions below the grey surface, i.e., inequality 
  (\ref{plotdef:ddXi}). The left plot shows qualitatively that the stability regions 
  decrease significantly for increasing values of $x$, in the relevant range of $0<y<1$. 
  From the right plot, one verifies that that the stability regions decrease significantly for 
  decreasing values of $y$. See the text for more details.}
  \label{dilaton-blackhole2}
\end{figure}

Consider as a first example the case for $\beta=1/2$, so that the stability regions, governed 
by the inequalities (\ref{stable_ddms}) and (\ref{plotdef:ddXi}), are depicted in Fig. 
\ref{dilaton-blackhole1}. The left plot describes the stability regions above the grey 
surface, given by (\ref{stable_ddms}), while the right plot describes the stability regions 
below the grey surface, i.e., inequality (\ref{plotdef:ddXi}). Note that the stability condition dictated by the inequality (\ref{stable_ddms}) shows that the stability regions decrease significantly for increasing values of $x$, within the considered range of $y$, i.e., $0<y<1$. In counterpart, inequality (\ref{plotdef:ddXi}) shows that the stability regions decrease significantly for decreasing values of $y$. These latter stability regions further decrease for increasing values of $x$.
As a second case consider the value $\beta=1/8$, depicted in Fig. \ref{dilaton-blackhole2}. We verify that the qualitative results are similar to the specific case of $\beta=1/2$, considered above.

\begin{figure}[!htb]
  %\centering
  \includegraphics[width=3.5 in]{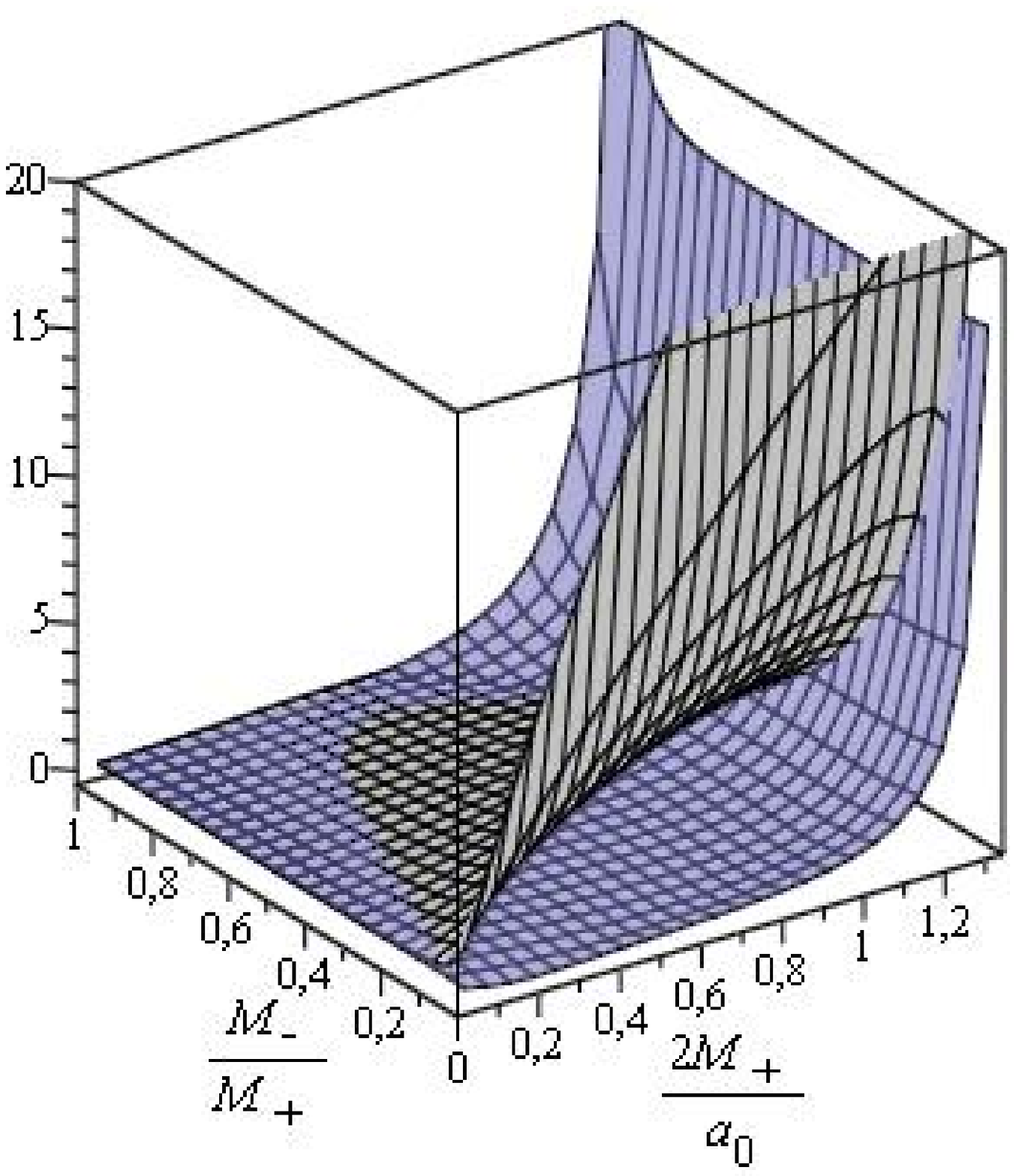}
  \includegraphics[width=3.5 in]{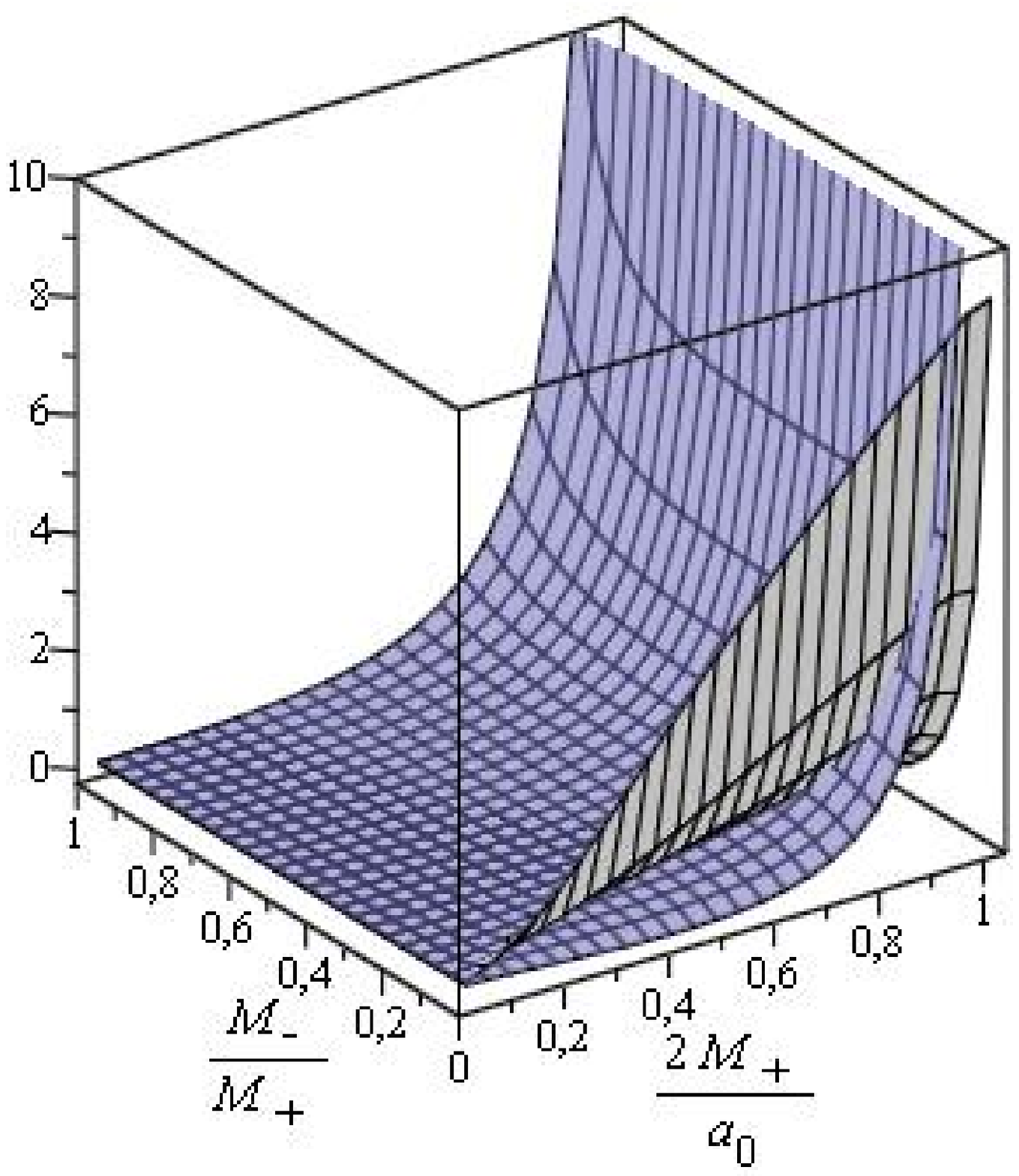}
  \caption{The plots depict the final stability regions for the thin shell dilaton wormhole. 
  The left plot describes the stability regions given by $\beta=1/2$, while the right plot
  describes stability regions given by $\beta=1/8$. 
  The final stability regions are depicted in between the surfaces of different shades of 
  grey. Note that the stability regions decrease for decreasing values of $\beta$. More 
  specifically, for decreasing values of $\beta$, stability regions exist for practically very 
  low values of $y$, i.e., $M_- \leq M_+$. In this context, it is interesting to note that in 
  the vicinity of the event horizon, the stability regions increase for increasing values of 
  $\beta$, provided one has low values of $y$. See the text for more details.
  }
  \label{dilaton-blackhole3}
\end{figure}

Collecting the above results, the final stability regions, depicted in between the surfaces of different shades of grey, are given in Fig. \ref{dilaton-blackhole3}. The left and right plot are given for the values of $\beta=1/2$ and $\beta=1/8$, respectively.
Note that the stability regions decrease for decreasing values of $\beta$. More specifically, for decreasing values of $\beta$, stability regions exist for practically very low values of $y$, i.e., $M_- \leq M_+$. In this context, it is interesting to note that in the vicinity of the event horizon, the stability regions increase for increasing values of $\beta$, provided one has low values of $y$.

%%%%%%%%%%%%%%%%%%%%%%%%%%%%%%%%%%%%%%%%%%%%%%%%%%%%%%%%%%%%%%%%%%%
\section{Discussion and Conclusion}\label{conclusion}
%%%%%%%%%%%%%%%%%%%%%%%%%%%%%%%%%%%%%%%%%%%%%%%%%%%%%%%%%%%%%%%%%%%

In this work, we have developed an extremely general, flexible and 
robust framework, leading to the linearized stability analysis of spherically symmetric thin shells. The analysis is well-adapted to general spherically symmetric thin shell traversable wormholes and, in this context, the construction confines the exotic material to the thin-shell. The latter, while constrained by spherical symmetry is allowed to move freely within the bulk spacetimes, which permits a fully dynamic analysis. To this effect, we have considered in great detail the presence of a flux term, which has been widely ignored in the literature. This flux term corresponds to the net discontinuity in the conservation law of the surface stresses of the bulk momentum flux, and is physically interpreted as the work done by external forces on the thin shell.

Relative to the linearized stability analysis, we have reversed the logic flow typically considered in the literature, and introduced a novel approach to the analysis. We recall that the standard procedure extensively used in the literature is to define a parametrization of the stability of equilibrium, so as not to specify an equation of state on the boundary surface \cite{Poisson,Eiroa,LoboCraw}. More specifically, the parameter $\eta(\sigma)= d{\cal P}/d\sigma$ is usually defined, and the standard physical interpretation of $\eta$ is that of the speed of sound. In this work, rather than adopt the latter approach, we  
considered that the stability of the wormhole is fundamentally linked to the behaviour of the surface mass $m_s(a)$ of the thin shell of exotic matter, residing on the wormhole throat, via a pair of stability inequalities. More specifically, we have considered the surface mass as a function of the potential. This novel procedure implicitly makes demands on the equation of state of the matter residing on the transition layer, and demonstrates in full generality that the stability of thin shell wormholes is equivalent to choosing suitable properties for the material
residing on the thin shell.

%---------------------

We have applied the latter stability formalism to a number of specific examples of particular importance: some presented to emphasize the features specific to possible asymmetry between the two universes used in traversable wormhole construction; some to emphasize the importance of NEC non-violation in the bulk; and some to assess the simplifications due to symmetry between the two asymptotic regions. In particular, we have considered the case of borderline NEC non-violation in the bulk. This is motivated by the knowledge that, on extremely general grounds, the NEC must be violated somewhere in the spacetime of a generic traversable wormhole, and if this were to happen in the bulk region, this would be equivalent to imposing $\Phi'_\pm(r)< 0$ in the bulk.  Thus to minimize NEC violations for thin-shell wormholes, we have considered the specific case of $\Phi_\pm=0$, which is particularly interesting for its mathematical simplicity, and for its physical interest as it corresponds to the constraint that the bulk regions on either side of the wormhole throat be on the verge of violating the NEC. 
We have also considered the simplification when the two bulk regions are identical, and analysed the stability regions of asymmetric thin-shell Schwarzschild wormholes and thin-shell Reissner-Nordstr\"{o}m wormholes in great detail. 
It was instructive to consider explicit cases which violate some of the energy conditions in the bulk. For instance, we considered thin-shell variants of the Ellis wormhole and two new toy models, and explored the linearized stability analysis in the presence of zero momentum flux and non-zero external forces. Finally, we analyzed thin-shell dilatonic wormholes, where the exterior spacetime solutions corresponded to a combined gravitational-electromagnetic-dilaton system.

%------------------------------------

In conclusion, by considering the matching of two generic static spherically symmetric spacetimes using the cut-and-paste
procedure, we have analyzed the stability of spherically symmetric dynamic thin-shell traversable wormholes --- stability to linearized
spherically symmetric perturbations around static solutions. The analysis provides a \emph{general and unified framework} for simultaneously addressing a large number of wormhole models scattered throughout the literature.  As such we hope it will serve to bring some cohesion and focus to what is otherwise a rather disorganized and disparate collection of results. A key feature of the current analysis is that we have been able to include ``external forces'' in the form of non-zero values for the metric functions $\Phi_\pm(r)$. (This feature is absent in much of the extant literature.) Another key aspect of the current analysis is the focus on $m_s(a)$, the ``mass'' of the thin shell of exotic matter residing on the wormhole throat --- and the realization that stability of the wormhole is fundamentally linked to the behaviour of this exotic matter via a pair of simple and relatively tractable inequalities.

%%%%%%%%%%%%%%%%%%%%%%%%%%%%%%%%%%%%%%%%%%%%%%%%%%%%%%%%%%%%%%%%%%%%
\acknowledgments
%%%%%%%%%%%%%%%%%%%%%%%%%%%%%%%%%%%%%%%%%%%%%%%%%%%%%%%%%%%%%%%%%%%%

N.M.G. acknowledges financial support from CONACYT-Mexico. F.S.N.L. acknowledges financial support of the
Funda\c{c}\~{a}o para a Ci\^{e}ncia e Tecnologia through Grants PTDC/FIS/102742/2008,
CERN/FP/116398/2010, CERN/FP/123615/2011 and CERN/FP/123618/2011.
M.V. acknowledges financial support from the Marsden Fund, and from a James Cook Fellowship, 
both administered by the Royal Society of New Zealand.
The authors wish to thank Prado Martin--Moruno for comments and feedback.

\enlargethispage{50pt}

%\clearpage

%%%%%%%%%%%%%%%%%%%%%%%%%%%%%%%%%%%%%%%%%%%%%%%%%%%%%%%%%%%%%%%%%%%%

%%%%%%%%%%%%%%%%%%%%%%%%%%%%%%%%%%%%%%%%%%%%%%%%%%%%%%%%%%%%%%%%%%%%
\end{document}